# Bandwidth Efficient Livestreaming in Mobile Wireless Networks: A Peer-to-Peer ACIDE Solution

Andrei Negulescu, Weijia Shang

*Abstract*— **In mobile wireless networks, livestreaming in high user density areas presents two typical challenges: the wireless bandwidth is depleted and the number of users is limited. In this study, a media distribution model utilizing peer–to-peer communications, Active Control in an Intelligent and Distributed Environment, is proposed for bandwidth efficient livestreaming. The basic idea is to group users with identical livestream interest in a cluster of *n* peers. Instead of sending *n* copies of a livestream package, only one copy is sent to the cluster. A package is divided into *n* blocks. Each user receives one block from the base station and the remaining *n*-1 blocks from the other peers. Two optimization problems are addressed. The first problem is minimizing the bandwidth needed to guarantee a continuous live media play on all peers. A solution is proposed to find the optimal block sizes such that the wireless bandwidth is minimized. The second problem is maximizing the number of peers admitted to a cluster, given a fixed wireless bandwidth. This problem is NP-complete and a greedy strategy is proposed to calculate a feasible solution for peer selection. The proposed model improves the bandwidth efficiency and allows more users to be served.**

## I. Introduction

Media livestreaming is among the most popular mobile wireless network data services. In high density areas (e.g. airplanes, trains, educational institutions, and sport venues), if the interest in a same livestream media becomes large, the bandwidth of the base station sending the livestream can be precipitously exhausted. Also, the bandwidth available for each data user is rapidly reduced. Consequently, the *network capacity*, the total number of users able to livestream the same media, is limited. In this study, an original model and solutions using *Peer-to-Peer* (P2P) communications are proposed to minimize the bandwidth and maximize the network capacity.

### A. Basic Idea

To improve the livestreaming bandwidth efficiency, an *Active Control in an Intelligent and Distributed Environment* (ACIDE) media distribution model is proposed. The essential components are a *base station* and a *cluster* formed by *n* users livestreaming the same media. Users admitted to a cluster are configured as *peers* able to establish P2P communications in a radio local area network, without using base station bandwidth. The basic idea of the ACIDE model is to send the livestream media in *packages*. Instead of sending *n* copies of the requested media package to the *n* peers of the cluster, the media package is divided into *n blocks* and the base station sends one block to each peer. Then the peers send their blocks to the other *n* - 1 peers in the cluster. In other words, only one copy of the media is sent, therefore the required bandwidth can be reduced *n* times. The *allocated bandwidth*, the amount of bandwidth that a base station has to allocate to a cluster such that all peers play livestream media without interruptions, is a function of many parameters such as block sizes, and peers download and upload bandwidth. The question is: what should be the sizes of these blocks and what amount of bandwidth should be allocated for sending a block to a peer?

In this study two problems are formulated as optimization problems. The first problem is to find the optimal block sizes, that may be different, and the bandwidth for each peer in order to minimize the allocated bandwidth. The second problem is to find the maximum number of peers *n* that can be admitted to a cluster with the *reserved bandwidth*, a fixed amount of bandwidth given by the base station. The case where users leave and join a cluster during livestreaming is also discussed.

### B. Motivation

The more ubiquitous the media livestreaming in a mobile wireless network, the lower the wireless bandwidth available to the data services users. Furthermore, in highly dense areas more users may not be able to access wireless data services. For example, in August 2023, a group of robotaxis autonomous vehicles stopped in the middle of a San Francisco street. At the same time, about four miles away, a large number of people attending a music festival were accessing wireless network data services. Because of a lack of available wireless bandwidth, the robotaxis, that consume a significant amount of bandwidth, were unable to receive route instructions from a remote operator [1], [2]. Our proposed ACIDE model could have addressed this situation by grouping the users livestreaming the same media at the festival into a cluster and by reserving the minimum bandwidth for a cluster formed with the robotaxis.

### C. Contributions

Before we present our main contributions, the following discussions are needed. The *livestream ratio* is the bandwidth used by a base station to distribute live media to one peer. Then, in a unicast model, if *n* peers livestream the same media, the allocated bandwidth is *n* times the livestream ratio. A more efficient communication model for sending a package to *n* peers is broadcast, where the allocated bandwidth equals the livestream ratio. However, broadcast is not ideal for livestreaming media with restricted access. Additionally, its bandwidth efficiency is reduced when a radio channel is shared with unicast traffic. Multicast is a better solution for media livestreaming because one package can be sent in one operation to a cluster of peers. The *multicast bandwidth* is equal to the livestream ratio. Given the challenges involved with its direct implementation in a wireless network, multicast is typically built on top of either broadcast or unicast communication layers. With multicast over broadcast, all peers download media at a minimum bandwidth equal to the livestream ratio. Peers with wireless bandwidth falling below this threshold may not receive media packages. Other peers with more bandwidth resources might not be able to access a higher-quality live media. If multicast over unicast is used, then $n$ equal sized packages are distributed to $n$ peers with a bandwidth equal to $n$ times the livestream ratio. The ACIDE communication model is a collaborative approach closer to the multicast definition, meaning that one package is sent to $n$ peers in one operation. This operation consists of sending the $n$ blocks of a media package to $n$ peers and exchanging the blocks between peers. More discussions on related work are presented in section VII.

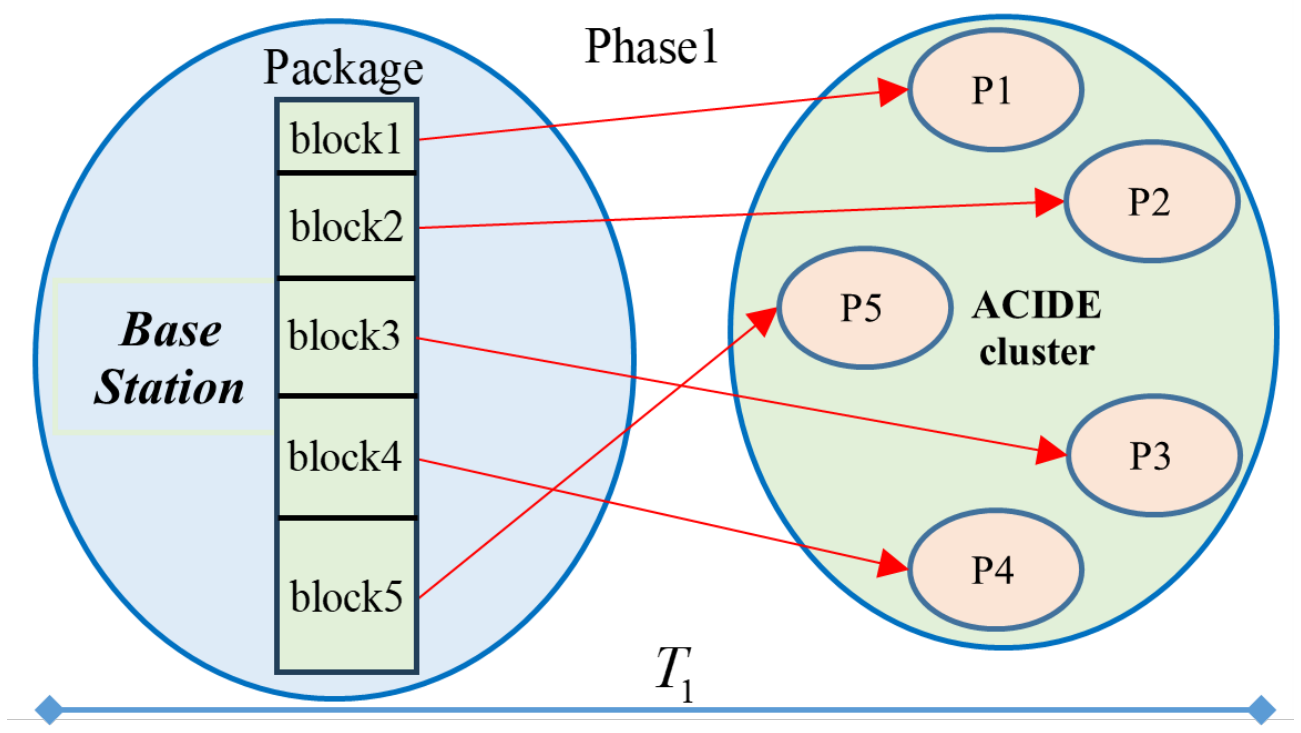

Fig. 1. ACIDE P2P cluster configured for Phase 1 when $n$=5

The major contributions in this study are summarized as follows. We present a low complexity method for dividing a live media package into optimal blocks such that the minimum amount of base station bandwidth is needed for sending all the media blocks to peers. An important result is that for a very large number of peers and an increasing average peer upload bandwidth the minimum allocated bandwidth is getting closer to the multicast bandwidth. We propose a feasible solution to an NP-Complete problem of maximizing the number of peers for a reserved bandwidth, such that a continuous livestream media play is guaranteed. The proposed greedy strategy guarantees that the largest amount of reserved bandwidth is allocated to the maximum number of peers admitted to a cluster. We also propose an efficient solution to handle the case where users can leave or join a cluster during the livestreaming.

### D. Organization

The rest of the paper is organized as follows. In section II models, definitions, and assumptions are presented. Sections III and IV present the bandwidth minimization problem and the network capacity maximization problem, respectively. Section V describes a dynamic case where users leave or join a cluster. Simulation results and related work are discussed in sections VI and VII respectively. Section VIII concludes the paper.

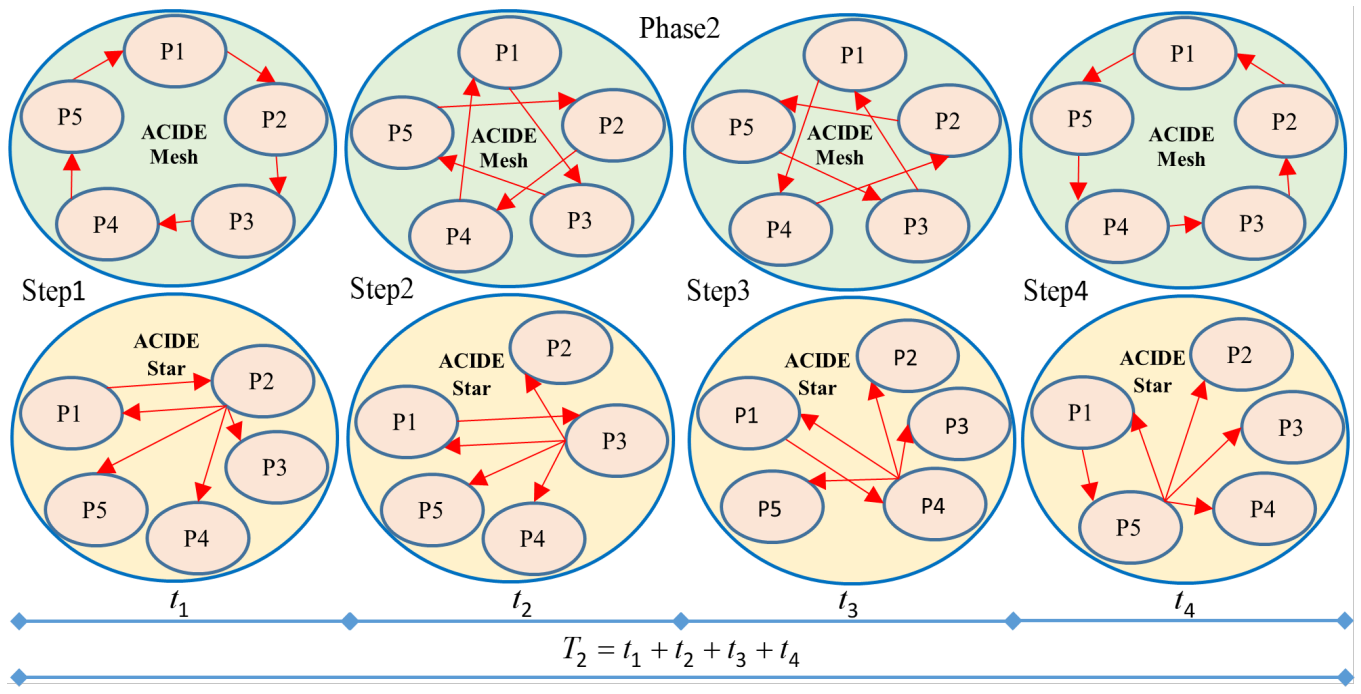

Fig. 2. ACIDE P2P configurations for the steps of Phase 2 when $n$=5

## II. The ACIDE Peer-to-Peer Model

In this section, the ACIDE P2P model along with its definitions and assumptions is presented. A notation list is given in Appendix B.

### A. The ACIDE P2P Communication Model

The ACIDE P2P model consists of a base station and a cluster. As illustrated in Fig. 1, a number of users livestreaming the same media from a base station are grouped together in a cluster. Inside a cluster users become peers. The peers can directly communicate to each other without using base station bandwidth. A livestream is divided in packages. Each package is divided in blocks and the blocks are distributed to peers.

A cluster is formed based on the following three properties. First, all peers of a cluster have the ***interest property***, that is they request the same livestream media from the base station. Second, all peers of a cluster have the ***proximity property***, which means they are present in the coverage area of the base station and close to each other such that radio communications can be established (i.e., a peer can communicate directly with all other peers in the cluster). Third, a peer has the ***resource property***, meaning that a peer uses two radio interfaces for P2P unidirectional connections, one for download and the other for upload. The download interface is used for receiving blocks from the base station and other peers. The upload interface is used for sending its own block to the other peers in the cluster. The upload interface may send the block using unicast or broadcast communication models. The two interfaces can be reconfigured to provide any peer to any peer connectivity.

The interconnect configurations of a cluster are based on the mesh and star topologies presented in Fig. 2. There are $n$ peers in a given cluster. In an ACIDE mesh, any two peers can be connected directly by reconfiguring the interfaces. An ACIDE star can be reconfigured such that each peer (e.g. P5 in Step 4) can broadcast to all other $n$ - 1 peers on the upload interface and at the same time receive from only one peer (e.g. P1 in Step 4) on the download interface.

### B. Definitions and Assumptions

**Definition 1**: The *delay bound* $T$ is the time interval that guarantees a continuous media playback, that is if a peer receives an entire package within a delay less than or equal to $T$ the peer can play the media without interruptions.

**Definition 2**: A *media package of size S* is the live media that is distributed to a peer within a delay bound *T*. Then, using Definitions 1 and 2, the livestream ratio is equal to $\frac{S}{T}$.

For further discussions, the following assumptions are needed.

**Assumption 1**: The *download bandwidth* $d_i$ and the *upload bandwidth* $u_i$ represent the maximum bandwidth values that a peer *i* download and upload interfaces respectively can sustain throughout a delay bound *T*. In order to reduce the energy consumption of each peer it is assumed that $u_i \le d_i$.

We assume that a base station sends the same livestream media to all peers in parallel. There are two scenarios of distributing a package to the *n* peers of a cluster. In the first scenario, during *T* the base station sends each peer a copy of an entire media package of size *S*. In this scenario, the bandwidth the base station has to allocate to the cluster is $n\frac{S}{T}$. Therefore, a large bandwidth is required for livestreaming.

In the second scenario, a package is divided into *n* blocks with sizes $s_1,...,s_i,...,s_n$, where $\sum_{i=1}^{n} s_i = S$. The base station sends block *i* to peer *i*, as illustrated in Fig. 1. Then peer *i* distributes block *i* to all other peers in the cluster. At the same time, peer *i* receives the other *n* - 1 blocks from the other peers. In this scenario, the bandwidth the base station has to allocate to the cluster is proportional to $\frac{S}{T}$ because only one package is sent. The basic idea of the second scenario is used in this study.

Because in the second scenario the upload interface of each peer is reconnected to the download interfaces of all other *n* - 1 peers as shown in Fig. 2, we assume the following:

**Assumption 2**: It is assumed that $u_i \le d_j$ for $i = 1,...,n$, $j = 1,...,n$.

### C. Problems

The problems addressed in this study are formulated below.

**Problem 1**: Given a cluster of *n* peers with download and upload bandwidth $d_i$ and $u_i$, and the delay bound *T*, how should a package be divided into *n* blocks with sizes $s_1,...,s_i,...,s_n$ and what is the value of the bandwidth $bw_i$, $i = 1,...,n$ allocated to each peer such that the total bandwidth required $\sum_{i=1}^{n} bw_i$ is minimized? In other words, given parameters $d_i$ and $u_i$, $i = 1,...,n$, *S* and *T*, what values should be selected for $s_i$ and $bw_i$, $i = 1,...,n$, to minimize $\sum_{i=1}^{n} bw_i$?

**Problem 2**: Given a reserved bandwidth *BW*, a number of users *N* with download and upload bandwidth $d_j$ and $u_j$, $j = 1,...,N$, and the delay bound *T*, what is the maximum number of users $n \le N$ that can be admitted as cluster peers, how should a package be divided into *n* blocks with sizes $s_1,...,s_i,...,s_n$, and what is the bandwidth value $bw_i, i = 1,...,n$ allocated to each peer such that $\sum_{i=1}^{n} bw_i$ is minimized and $\sum_{i=1}^{n} bw_i \le BW$? In other words, given parameters $N$, $d_j$ and $u_j$, $j = 1,...,N$, *S* and *T*, how should the reserved bandwidth *BW* be divided and allocated to a maximum number of $n \le N$ peers and what values should be selected for $s_i$ and $bw_i$, $i = 1,...,n$, such that a continuous media playback on all peers is guaranteed for a minimum $\sum_{i=1}^{n} bw_i \le BW$?

### D. Procedures

A media package distribution to the peers of an ACIDE cluster is performed in two phases. In Phase 1, the base station divides a package into *n* blocks and then allocates *bandwidth* $bw_i$ to send block *i* to peer *i*, where $i = 1,...,n$. The time for Phase 1 is denoted as $T_1$. Let $B(i)$ denote the event that the base station sends block *i* to peer *i*.

In Phase 2, peer *i* sends block *i* to peer *j* and receives block *j*, where $j = 1,...,i-1,i+1,...,n$, from the other $n$ - 1 peers. In Section III it is proven that the most bandwidth efficient solution is reached if Phase 2 starts after every peer receives its block in Phase 1. Therefore, the two phases are sequential and they don't overlap. Phase 2 uses multiple *point-to-point* or *multipoint* [3] group communications that allow the peers to establish two P2P concurrent sessions: one for upload and one for download. The P2P communications in Phase 2 do not use base station bandwidth. The time for Phase 2 is denoted as $T_2$. In general, it takes *n* - 1 steps for each peer to receive all *n* - 1 blocks from the other peers.

Let $M(i, j)$ and $R(i, j)$ denote the event that peer *i* sends its block to peer *j* in the mesh and star cluster respectively, $i = 1,...,n$, $i \ne j$. The following procedures describe how blocks are distributed to the peers of a mesh and a star cluster. It is important to highlight that all events in the same row take place in parallel and the steps are sequential. Then, we have:

**Procedure 1 (Mesh)**:

**Phase 1**: $B(1)$, $B(2)$, $B(3)$,...,$B(n-1)$, $B(n)$

**Phase 2**:

**Step 1**: $M(1,2)$, $M(2,3)$, $M(3,4)$,...,$M(n\text{-}1,n)$, $M(n,1)$

**Step 2**: $M(1,3)$, $M(2,4)$, $M(3,5)$,...,$M(n\text{-}1,1)$, $M(n,2)$

...

**Step (*n*-1)**: $M(1,n)$, $M(2,1)$, $M(3,2)$,...,$M(n\text{-}1,n\text{-}2)$, $M(n,n\text{-}1)$

**Procedure 2 (Star)**:

**Phase 1:** $B(1),\ B(2),\ B(3),...,B(n-1),\ B(n)$

**Phase 2**:

**Step 1**: $R(1,2),\ R(2,1),\ R(2,3),...,R(2,n-1),\ R(2,n)$

**Step 2:** $R(1,3),\ R(3,1),\ R(3,2),...,R(3,n-1),\ R(3,n)$

...

**Step (*n*-1):** $R(1,n),\ R(n,1),\ R(n,2),...,R(n,n-2),\ R(n,n-1)$

More discussions on how the *n* blocks of a package are distributed to *n* peers are presented next. The blocks are sent from the base station to the peers in parallel during Phase 1. Because a package is divided into *n* blocks, a block *i* is sent to peer *i* with a bandwidth $bw_i \le d_i,\ i=1,...,n$. In this case, the time of Phase 1 is $T_1 = \max\{\frac{s_i}{\min\{bw_i, d_i\}}\} = \frac{s_j}{bw_j}$, for a peer *j*, $1 \le j \le n$. Then, $bw_j = \frac{s_j}{T_1}$. Moreover, $\sum_{i=1}^{n} d_i \ge \sum_{i=1}^{n} bw_i \ge \frac{S}{T_1}$ and we assume the following:

**Assumption 3**: It is assumed that the sum of the download bandwidth of all peers has to satisfy: $\sum_{i=1}^{n} d_i \ge \frac{S}{T_1}$.

Without the above assumption, a media package of size *S* cannot be downloaded from the base station within time $T_1$ and therefore, the media cannot be played continuously by peers.

An example for *n* = 5 is shown in Fig. 1, where five peers livestream the same media from a base station. Throughout Phase 1 each peer is downloading a block from the base station in time $T_1$. Fig. 2 shows the $n-1$ steps of Phase 2. The mesh and star clusters are configured in order to implement all concurrent events taking place in each step. A peer reconnects its two unidirectional interfaces to different peers in each of the *n* - 1 steps. As a result, instead of using *n* - 1 simultaneous, bidirectional connections, a peer is using the equivalent of only one bidirectional connection to transfer blocks to and from another peer, at any time during Phase 2.

### *E. The Former P2P Communication Model*

The ACIDE model borrows some concepts from a previous P2P communications model proposed in [4], [5]. The model in [4], [5] is used to distribute pre-recorded media in a wireline network. The ACIDE model is used for media livestreaming in a mobile wireless network. In the wireline model, each peer requires *n* - 1 simultaneous, bidirectional connections for media transfers to other peers and $\frac{n(n-1)}{2}$ concurrent bidirectional connections are established. A peer in the ACIDE model is using only one bidirectional connection at a time to exchange media. Then, *n* unidirectional concurrent connections are needed for a mesh cluster and two for a star cluster. The goal of the wireline model is minimizing the media distribution time while the ACIDE model is minimizing the allocated bandwidth.

## III. Bandwidth Optimzation With $n$ Peers

In this section, Problem 1 is formulated as an optimization problem and a solution is proposed. Our goal is to minimize the allocated bandwidth $\sum_{i=1}^{n} bw_i$ and to allow the peers to play livestream media continuously. Problem 1 can be stated below:

$$\text{Minimize} \quad \sum_{i=1}^{n} bw_i$$

$$\text{Subject to} \quad T_1 + T_2 \le T$$

*T* is the delay bound. $T_1$ and $T_2$ are the times of Phase 1 and Phase 2, respectively. Both $T_1$ and $T_2$ are functions of parameters: $s_i$, $bw_i$, $d_i$, and $u_i$, $i=1,...,n$. In Phase 1, the base station sends block *i* to peer *i* with the allocated bandwidth $bw_i$. Hence, the time that it takes for this operation to complete is $\frac{s_i}{bw_i}$, where $bw_i \le d_i$. Therefore:

$$T_1 = \max\{\frac{s_i}{bw_i}, i=1,....,n\} \tag{1}$$

The block distribution Procedure 1 and Procedure 2 are used to determine the times of all steps in Phase 2. Let $t_i$ denote the time for step *i*, $i=1,...,n-1$. Then, the total time of Phase 2 is:

$$T_2 = \sum_{i=1}^{n-1} t_i \tag{2}$$

For the mesh topology, the times of all steps in Phase 2 are:

$$\begin{gathered} t_1 = \max\{\frac{s_1}{\min\{u_1,d_2\}}, \frac{s_2}{\min\{u_2,d_3\}}, ...., \frac{s_n}{\min\{u_n,d_1\}}\} \\ ... \\ t_{n-1} = \max\{\frac{s_1}{\min\{u_1,d_n\}}, \frac{s_2}{\min\{u_2,d_1\}}, ...., \frac{s_n}{\min\{u_n,d_{n-1}\}}\} \end{gathered} \tag{3}$$

For the star topology, the finish times of all steps are:

$$\begin{gathered} t_1 = \max\{\frac{s_1}{\min\{u_1,d_2\}}, \frac{s_2}{\min\{u_2,d_1\}}, ...., \frac{s_2}{\min\{u_2,d_n\}}\} \\ ... \\ t_{n-1} = \max\{\frac{s_1}{\min\{u_1,d_n\}}, \frac{s_n}{\min\{u_n,d_1\}}, ...., \frac{s_n}{\min\{u_n,d_{n-1}\}}\} \end{gathered} \tag{4}$$

**Theorem 1**: The objective function $\sum_{i=1}^{n} bw_i$ is minimized if and only if all events in Phase 1 take the same time and all the events take the same time in Phase 2, during each step. That is, in Phase 1, $\frac{s_i}{bw_i} = \frac{s_j}{bw_j}$ for $i=1,...,n$, $j=1,...,n$, and in Phase 2, the events of a step *i*, $i=1,...,n-1$, in (3) and (4) are equal.

The proof of Theorem 1 is presented in Appendix A.

**Lemma 1**: The time of Phase 1 is:

$$T_1 = \frac{s_1}{bw_1} = \frac{s_2}{bw_2} = \ldots \frac{s_n}{bw_n} \tag{5}$$

**Proof**: From Theorem 1, in the optimal case, the times of all events $B(i)$ in Phase 1 are equal. Then, from (1) we get (5). □

From (5) and Theorem 1, Assumption 3 is verified because $\sum_{i=1}^{n} bw_i = \frac{S}{T_1}$ is the minimum allocated bandwidth necessary to transfer all the blocks of a package of size $S$ in time $T_1$ and to allow an uninterrupted livestream media play by the peers.

Furthermore, $\min\{u_i, d_j\} = u_i$ because of Assumption 2. Then, for the mesh topology in Phase 2 we have the following:

$$\begin{aligned} &t_1 = \frac{s_1}{\min\{u_1, d_2\}} = \ldots = \frac{s_n}{\min\{u_n, d_1\}} = \frac{s_1}{u_1} = \frac{s_2}{u_2} = \ldots = \frac{s_n}{u_n} \\ &\ldots \\ &t_{n-1} = \frac{s_1}{\min\{u_1, d_n\}} = \ldots = \frac{s_n}{\min\{u_n, d_{n-1}\}} = \frac{s_1}{u_1} = \frac{s_2}{u_2} = \ldots = \frac{s_n}{u_n} \end{aligned} \tag{6}$$

Similarly, for the star topology in Phase 2:

$$\begin{aligned} &t_1 = \frac{s_1}{\min\{u_1, d_2\}} = \ldots = \frac{s_2}{\min\{u_2, d_n\}} = \frac{s_1}{u_1} = \frac{s_2}{u_2} = \ldots = \frac{s_2}{u_2} \\ &\ldots \\ &t_{n-1} = \frac{s_1}{\min\{u_1, d_n\}} = \ldots = \frac{s_n}{\min\{u_n, d_{n-1}\}} = \frac{s_1}{u_1} = \frac{s_n}{u_n} = \ldots = \frac{s_n}{u_n} \end{aligned} \tag{7}$$

Then, in Phase 2, according to (6) and (7), $t_1 = \ldots = t_{n-1}$ and:

$$\frac{s_1}{u_1} = \frac{s_2}{u_2} = \ldots = \frac{s_i}{u_i} = \ldots = \frac{s_n}{u_n} \tag{8}$$

**Lemma 2**: The time of Phase 2 for both topologies is:

$$T_2 = (n-1)\frac{s_1}{u_1} = (n-1)\frac{s_2}{u_2} = \ldots = (n-1)\frac{s_n}{u_n} \tag{9}$$

**Proof**: From Theorem 1, Phase 2 does not start until Phase 1 is complete. Because in each step the events $M(i,j)$, $R(i,j)$ take the same time, step $i$ does not start until all the events in step $i$ - 1 are completed. Then, from (8) the times of all steps in Phase 2 are equal $t_1 = \ldots = t_{n-1} = \frac{s_1}{u_1} = \frac{s_i}{u_i}$ and $T_2 = \sum_{i=1}^{n-1} t_i = (n-1)\frac{s_i}{u_i}$.□

To solve Problem 1, the basic idea is to find the optimal block sizes $s_i$ first and then calculate $T_2, T_1$ and find $bw_i = \frac{s_i}{T_1}$, $i = 1,\ldots,n$. We discuss the optimal sizes $s_i$, $i = 1,\ldots,n$ next.

**Lemma 3**: The optimal values of $s_i$ are given by the equations:

$$\begin{aligned} &\frac{s_k}{u_k}\sum_{i=1}^{k} u_i + \sum_{i=k+1}^{n} s_i = S, k = 1,\ldots,n \\ &k = 1: \quad s_1 + \sum_{i=2}^{n} s_i = S \\ &\ldots \\ &k = n: \quad \frac{s_n}{u_n}\sum_{i=1}^{n} u_i = S \end{aligned} \tag{10}$$

**Proof**: According to (8) and from $\sum_{i=1}^{n} s_i = S$, we can present $n$ equalities in the following matrix notation:

$$\begin{bmatrix} 1 & 1 & \ldots & 1 & 1 \\ \frac{1}{u_1} & \frac{-1}{u_2} & \ldots & 0 & 0 \\ \ldots & \ldots & \ldots & \ldots & \ldots \\ 0 & 0 & \ldots & \frac{1}{u_{n-1}} & \frac{-1}{u_n} \end{bmatrix} \cdot \begin{bmatrix} s_1 \\ s_2 \\ \ldots \\ s_{n-1} \\ s_n \end{bmatrix} == \begin{bmatrix} S \\ 0 \\ \ldots \\ 0 \end{bmatrix}$$

From the above we have $s_1 = \frac{u_1}{u_2}s_2$ and $\frac{u_1}{u_2}s_2 + s_2 + \sum_{i=3}^{n} s_i = S$. Then $\frac{u_1 + u_2}{u_2}s_2 + \sum_{i=3}^{n} s_i = S$. Similarly, we have $s_2 = \frac{u_2}{u_3}s_3$ and $\frac{u_1 + u_2}{u_2}s_2 + s_3 + \sum_{i=4}^{n} s_i = S$, then $\frac{u_1 + u_2}{u_2}\frac{u_2}{u_3}s_3 + s_3 + \sum_{i=4}^{n} s_i = S$ and $\frac{u_1 + u_2 + u_3}{u_3}s_3 + \sum_{i=4}^{n} s_i = S$. In general, we get (10). From (10) we calculate the optimal values of $s_i$. □

**Definition 3**: Let $\alpha_k = \frac{1}{u_k}\sum_{i=1}^{k} u_i$, $k = 2,\ldots,n$, $\alpha_1 = 0$ if $k = 1$.

Using Definition 3, all the above $n$ equations in (10) can be expressed in the following matrix notation in (11), where $\mathbf{s} = [s_1,\ldots,s_n]^t$ is the solution vector. The optimal sizes $s_i$, $i = 1,\ldots,n$, are calculated from the system of linear equations in (11). The optimal solution $\mathbf{s} = [s_1,\ldots,s_n]^t$ with Lemmas 1 and 2 are used to derive an optimal solution for $bw_i$ next.

$$\begin{bmatrix} 1 & 1 & \ldots & 1 & 1 \\ 0 & \alpha_2 & \ldots & 1 & 1 \\ \ldots & \ldots & \ldots & \ldots & \ldots \\ 0 & 0 & \ldots & \alpha_{n-1} & 1 \\ 0 & 0 & \ldots & 0 & \alpha_n \end{bmatrix} \cdot \begin{bmatrix} s_1 \\ s_2 \\ \ldots \\ s_{n-1} \\ s_n \end{bmatrix} = \begin{bmatrix} S \\ S \\ \ldots \\ S \\ S \end{bmatrix} \tag{11}$$

**Lemma 4**: The minimum allocated bandwidth $bw_i$ to peer *i* is:

$$bw_i = \frac{s_i}{T-(n-1)\frac{s_i}{u_i}} \tag{12}$$

**Proof**: From (5) $bw_i = \frac{s_i}{T_1}$. Then, from (9) and because $T_1 = T - T_2 = T-(n-1)\frac{s_i}{u_i}$, we get the optimal $bw_i$ in (12). □

In Phase 1, according to (5), $bw_i = \frac{s_i}{s_1} bw_1$, $i = 1,...,n$, and $bw = \sum_{i=1}^{n} bw_i = \sum_{i=1}^{n} \frac{s_i}{s_1} bw_1 = \frac{bw_1}{s_1}\sum_{i=1}^{n} s_i = \frac{bw_1}{s_1} S$. Then, in general:

$$bw = \frac{bw_i}{s_i} S,\ i = 1,...,n \tag{13}$$

**Theorem 2:** If $s_i$, $i = 1,...,n$, are optimal then the values of $bw_i = \frac{s_i}{T_1}$ are minimum and $bw = \sum_{i=1}^{n} bw_i$ is minimum.

**Proof:** The proof is immediate from Lemmas 1, 2, and 3. □

Three observations from Theorems 1, 2 are discussed next.

**Observation 1**: If the ACIDE P2P communication model is not used, and *n* users are livestreaming the same media from a base station in parallel, the whole cluster would require a bandwidth $n\frac{S}{T}$ for the download of a package.

If the ACIDE P2P communication model is used, the *minimum allocated bandwidth bw* is calculated using (8) and (12) as follows. From Lemma 4 we have $bw = \sum_{i=1}^{n} bw_i = \frac{S}{T - T_2}$. According to (10), for $k = n$ we have $\frac{s_n}{u_n} = \frac{S}{\sum_{i=1}^{n} u_i}$. Then, the minimum allocated bandwidth should satisfy the following:

$$bw = \sum_{i=1}^{n} bw_i == \frac{S}{T-(n-1)\frac{s_n}{u_n}} = \frac{S}{T-(n-1)\frac{S}{\sum_{i=1}^{n} u_i}} \tag{14}$$

Let the *average upload bandwidth* of a cluster be $u_{avg} = \frac{1}{n}\sum_{i=1}^{n} u_i$. Then from (14) we have $T-(n-1)\frac{S}{\sum_{i=1}^{n} u_i} = T - \frac{n-1}{n}\frac{S}{u_{avg}}$ and

$$bw = \frac{S}{T - \frac{n-1}{n}\frac{S}{u_{avg}}} \tag{15}$$

From (15), three possible *bw* results are presented next. First, if many peers with $u_i > u_{avg}$ join a cluster and $u_{avg}$ increases, then $T_2$ decreases and *bw* approaches the multicast bandwidth. Second, *bw* approaches $\frac{S}{T}$ if $u_{avg}$ increases and no new peers join the cluster. Third, as the number of new peers joining a cluster is getting larger while $u_{avg}$ remains fixed, *bw* approaches a constant value equal to:

$$bw = \frac{S}{T - \frac{S}{u_{avg}}}. \tag{16}$$

**Observation 2**: The second observation is that the quality of the live media distributed to the peers of a cluster may be changing over the duration of a livestream. This means that, depending on bandwidth availability, *S*, the size of a package distributed within a constant time *T*, may decrease or increase.

Because $bw > 0$, from (14) we have $T > (n-1)\frac{S}{\sum_{i=1}^{n} u_i}$, and $\sum_{i=1}^{n} u_i > (n-1)\frac{S}{T}$. Moreover, because $n\frac{S}{T} \geq \sum_{i=1}^{n} bw_i$, from (14) $n\frac{S}{T} \geq \frac{\frac{S}{T}}{1-\frac{(n-1)}{\sum_{i=1}^{n} u_i}\frac{S}{T}}$, and $\frac{n-1}{n} \geq \frac{(n-1)}{\sum_{i=1}^{n} u_i}\frac{S}{T}$. Consequently, because *S* can be adjusted while *T* is constant, a variable livestream ratio is supported by the model. Therefore, in order to reach a better live media quality, the livestream ratio may be increased up to its upper bound, given by $\frac{S}{T} \leq \frac{1}{n}\sum_{i=1}^{n} u_i = u_{avg}$.

**Observation 3:** The minimum allocated bandwidth *bw*, as determined by (15), has been calculated under the assumption that the blocks of a package are sent in parallel during Phase 1. However, if the base station uses a serial communication mode to transmit the blocks sequentially, the minimum allocated bandwidth *bw* remains the same. Let $\theta_i$ be the time that it takes peer *i* to download a block of optimal size $s_i$ in the serial case. If $\sum_{i=1}^{n} \theta_i \leq T_1$ the media can be continuously played by peers.

**Lemma 5**: If in Phase 1 the blocks of a package with size *S* are sent to peers using a serial communication mode, the minimum allocated bandwidth required to download all blocks of optimal size $s_i$ in time $T_1$ is equal to *bw*, and:

$$\theta_i = \frac{s_i}{bw} \tag{17}$$

**Proof:** From Assumption 3 and (5), an uninterrupted live media

**Algorithm 1:** Bandwidth Optimization with $n$ Peers

**Input**: $n, S, T, d_i, u_i, i = 1,...,n$

**Output**: $s_i$, $bw_i$, $bw$

1: Calculate $\mathbf{s} = [s_1,...,s_n]^t$ using the linear system in (11)
2: Calculate $T_2$ using (9), and find $T_1$ from $T_1 = T - T_2$
3: Calculate $\mathbf{bw} = [bw_1,...,bw_n]^t$ according to (12)
4: Calculate the minimum allocated bandwidth $bw$ using (13)

play is guaranteed if a package $S$ is sent with $bw$, meaning that each block $s_i$ is received by a peer $i$ in $T_1$. Then $bw = \frac{S}{T_1}$, and we have the following: $\sum_{i=1}^{n}\theta_i \le T_1 = \frac{1}{bw}\sum_{i=1}^{n}s_i = \frac{s_1}{bw} + \frac{s_2}{bw} + ... + \frac{s_n}{bw}$. For a minimum $bw$ and optimal $s_i$, $\theta_i = \frac{s_i}{bw}$ and $\sum_{i=1}^{n}\theta_i = T_1$. □

The solution to Problem 1 is calculated by Algorithm 1 and it can be found in four steps. Let $\mathbf{s} = [s_1,...,s_n]^t$ and $\mathbf{bw} = [bw_1,...,bw_n]^t$. In step 1, the size vector $\mathbf{s}$ is calculated according to (11). In step 2, according to (9), the value of $T_2$ is calculated. $T_1$ can be calculated from the fact that $T_1 = T - T_2$. In step 3, the values of $bw_i$, $i = 1,...,n$ can be calculated with (12). Then, from (13), $bw$ can be calculated. The complexity of finding the solution can be analyzed as follows. Because the matrix in (11) is triangular, the time complexity of finding $\mathbf{s}$ is $\theta(n^2)$. In step 3, the time complexity of calculating $bw_i$ is $\theta(n)$. Therefore, the overall time complexity is $\theta(n^2)$.

A numerical example for the cluster in Fig. 1 and Fig. 2 is as follows. We assume a delay bound of $T = 200$ ms and the following peers upload bandwidth $u_i \in \{15,17,18,19,20\}$ kbps. Then $u_{avg} = 17.8$ kbps. A minimum download bandwidth of $d_i = 20$ kbps and a livestream ratio of 10 kbps are considered. Then, in Algorithm 1 step 1 we calculate the block sizes $\mathbf{s} = [337,382,404,426,449]^t$ bits. In step 2 we calculate $T_2 = 89$ ms and $T_1 = 111$ ms. Then in step 3 and in step 4 we find $\mathbf{bw} = [3.061,3.469,3.673,3.877,4.081]^t$ kbps, and the minimum allocated bandwidth $bw = 18.163$ kbps respectively. In this study the numerical results are the integer floor approximations of the values calculated by the ACIDE model.

## IV. Number of Peers Optimzation Knowing the Reserved Bandwidth $BW$

In this section, the problem of finding the maximum number of peers $n$ that can be grouped in an ACIDE P2P cluster knowing the reserved bandwidth $BW$ is formulated as an optimization problem. A greedy strategy is proposed to calculate a feasible solution such that the largest possible amount of $BW$ is shared among the peers admitted to the cluster [6].

**Algorithm 2:** Number of Peers Optimization for a fixed $BW$

**Input**: $N, S, T, BW, d_i, u_i, i = 1,...,N$

**Output**: $n$, $s_i$, $bw_i$, $bw$, $L$

1: **Initialize:** $n = N$
2: Create list $L$ with $n$ users
3: Calculate $bw$ using **Algorithm 1**
4: **while** $bw > BW$
5: Set $n = n - 1$
6: Remove the user with the lowest $u_i$, update $L$
7: Calculate $bw$ using **Algorithm 1**
8: **end while**

Let $N$ be a number of users having the ***interest***, ***proximity*** and ***resource*** properties. The minimum allocated bandwidth to a cluster of $n \le N$ peers is $\sum_{i=1}^{n} bw_i$. Problem 2 is stated next.

$$\begin{aligned} &\text{Maximize} && n \\ &\text{Subject to} && T_1 + T_2 \le T \\ & && \sum_{i=1}^{n} bw_i \le BW \\ & && n \le N \end{aligned}$$

Problem 2 can be reformulated as an equivalent bandwidth allocation problem, where $BW$ is divided among $n \le N$ peers. Because this formulation is equivalent to the NP-complete *SUBSET_SUM* problem [7], Problem 2 is also NP-Complete. As presented in Algorithm 2, we propose a greedy strategy for the selection of peers admitted to a cluster. The ACIDE model bandwidth optimization method is used to decide if livestreaming to a cluster of $N$ peers is possible for a given $BW$. If it is not possible, a greedy strategy, removing the user with the lowest $u_i$ from the list of users $L$, is proposed to calculate a feasible solution, in polynomial time, in several iterations [6].

**Theorem 3**: Let $bw > BW$ be the minimum allocated bandwidth calculated for $N$ peers. If after removing the user with $u_j = \min\{u_i, i = 1,....,N\}$, the updated minimum allocated bandwidth is $bw \le BW$, then $N = N - 1$ is the maximum number of peers using the largest possible amount of $BW$.

**Proof:** Let $u_1 = \min\{u_i, i = 1,..,N\}$, $u_N = \max\{u_i, i = 1,..,N\}$ and $u_N = u_1 + \lambda$, $\lambda > 0$. From (15), for a cluster of $N$ peers we have $bw = \frac{S}{T - T_2} = \frac{S}{T - \frac{N-1}{N}\frac{S}{u_{avg}}}$, where $u_{avg} = \frac{\sum_{i=1}^{N} u_i}{N}$.

*Case1*: The user with the upload bandwidth $u_1$ is removed and $u_{avg,1} = \frac{\sum_{i=2}^{N} u_i}{N-1}$. Moreover, $u_{avg,1} = \frac{u_2 + ... + u_N}{N-1} = \frac{Nu_{avg} - u_1}{N-1}$.

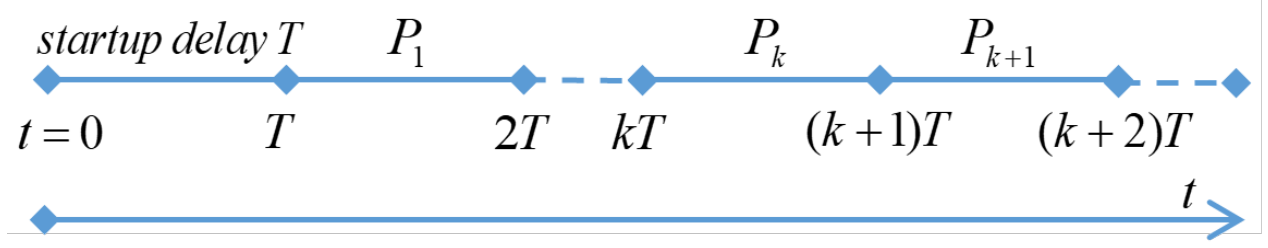

Fig. 3. Livestream media packages $P_1 ... P_{k+1} ... P_\infty$

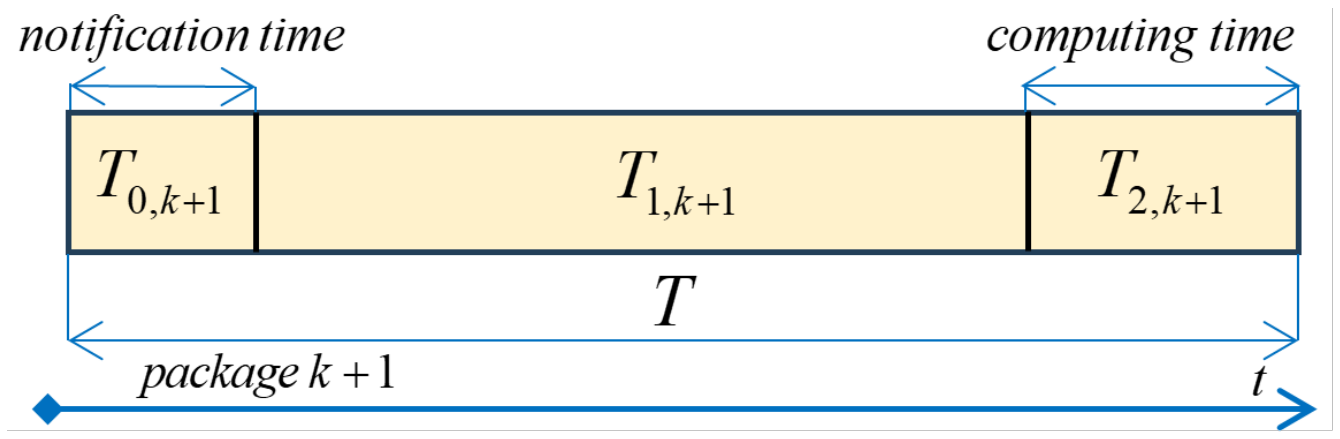

Fig. 4. Package $P_{k+1}$ distribution in the dynamic case

Then the updated minimum allocated bandwidth becomes $bw_{1,\min} = \dfrac{S}{T - \dfrac{(N-1)^2 S}{N(Nu_{avg} - u_1)}}$.

*Case2*: If the user with the upload bandwidth $u_N$ is removed $u_{avg,N} = \dfrac{\sum_{i=1}^{N-1} u_i}{N-1}$. Then $u_{avg,N} = \dfrac{Nu_{avg} - u_N}{N-1} = u_{avg,1} - \dfrac{\lambda}{N-1}$. Therefore, the updated minimum allocated bandwidth is $bw_{N,\max} = \dfrac{S}{T - \dfrac{(N-1)^2 S}{N(Nu_{avg} - u_1 - \lambda)}}$.

Because $Nu_{avg} - u_1 > Nu_{avg} - u_1 - \lambda$, the time $T_2$ of Case1 is less than the time $T_2$ of Case2. Clearly, $bw_{1,\min} < bw_{N,\max}$. If $bw_{1,\min} < BW$ and $bw_{N,\max} < BW$ the Algorithm 2 ends its execution and $BW - bw_{1,\min} > BW - bw_{N,\max}$. This means that $N = N-1$ is the maximum number of peers using the largest amount of reserved bandwidth *BW*. □

The maximum number of iterations of Algorithm 2 is *N*, meaning that at most *N* systems of linear equations given by (11) have to be solved. Because the complexity of calculating *bw* is $\theta(n^2)$ and $1 \le n \le N$, the solution to Problem 2 can be calculated with the overall time complexity of $\theta(n^3)$.

## V. Active Peer Control

In this section, an *Active Peer Control* (APC) method is proposed for a dynamic case where peers can leave or join a cluster and change their download and upload bandwidth due to varying conditions on radio propagation. How a livestream is divided in packages $P_k$, $k = 1,...,\infty$, is shown in Fig. 3. In the previous sections, where a static case is addressed, it is assumed that a cluster does not change, meaning that parameters *n*, $d_i$, and $u_i$ do not change for all the packages of a livestream. In the static case, two phases are the essentials to efficiently distribute the blocks of a package to peers. In the dynamic case shown in Fig. 4, we propose distributing a package $P_{k+1}$ in three phases: Phase 0, Phase 1, and Phase 2. The delay bound *T* is divided accordingly in times $T_{0,k+1}$, $T_{1,k+1}$ and $T_{2,k+1}$. It is assumed that in the dynamic case the parameters used to distribute package $P_k$ to *n* peers are saved in list $L_k = \{n, d_1, ..., d_n, u_1, ..., u_n\}$.

TABLE I. Active Peer Control Tasks During $P_{k+1}$ Distribution

| | **Phase 0** $T_{0,k+1}$ | **Phase 1** $T_{1,k+1}$ | **Phase 2** $T_{2,k+1}$ |
|---|---|---|---|
| **Base Station** | Sends notifications | Receives join or leave requests; Sends $P_{k+1}$ blocks $s_i$ to peers | Updates $L_{k+2}$; Calculates $s_i$, $bw_i$, $bw$ for $P_{k+2}$ |
| **Peers** | Receive notifications | Send join or leave requests; Receive $P_{k+1}$ blocks $s_i$ | Exchange $P_{k+1}$ blocks $s_i$ using P2P communications |
| **Base Station Bandwidth** | $bw_{0,k+1}$ | $bw_{1,k+1}$ | $bw_{2,k+1} = 0$ |

The basic idea of the dynamic case is as follows. To join or leave a cluster, peers send requests to the base station during $T_{1,k}$. Then, the base station updates $L_{k+1}$ and calculates $s_i$, $bw_i$ and $bw$ for $P_{k+1}$ in $T_{2,k}$. Notifications are sent to peers during $T_{0,k+1}$, the *notification time*, informing them about the cluster changes needed for the distribution of $P_{k+1}$. As illustrated in Fig. 4, peers start downloading the media blocks of $P_{k+1}$ after $T_{0,k+1}$. If parameters *n*, $d_i$, and $u_i$ do not change we have $T_{0,k+1} = 0$ and list $L_{k+1} = L_k$ is used to distribute $P_{k+1}$.

An outline of the tasks running on a base station and peers during the distribution of a package $P_{k+1}$ is given in Table I. If any parameter changes, the base station sends each peer a notification during $T_{0,k+1}$. As presented in Table I, during $T_{1,k+1}$ the base station sends the media blocks $s_i$ to the peers and during $T_{2,k+1}$ peers exchange media blocks $s_i$ using P2P communications. This study does not cover how the base station and peers identify bandwidth variations during $T_{1,k+1}$ and $T_{2,k+1}$, or how the base station detects when peers are leaving a cluster without sending requests. Bandwidth monitoring, and how peers initiate join and leave requests, are reserved for future work.

We propose that a notification includes two fields: a peer identification $1 \le pid \le n$ and the number of peers *n*. Both fields are used for Phase 2 communications. Each field is $\lceil log_2 n \rceil$ bits long. Then, a notification has $2\lceil log_2 n \rceil$ bits.

In $T_{0,k+1}$ bandwidth $bw_{0,k+1}$ is used to send *n* notifications in parallel, Then,

**Algorithm 3:** Active Peer Control

**Input** $S, T, L_k = \{N, d_1, ..., d_N, u_1, ..., u_N\}$

**Output** $s_i$, $bw_i$, $bw$, $L_{k+1} = \{n, d_1, ..., d_n, u_1, ..., u_n\}$

1: Update $L_{k+1} = \{n, d_1, ..., d_n, u_1, ..., u_n\}$ for $P_{k+1}$

2: Calculate $\mathbf{s} = [s_1, ..., s_n]^t$ with (11), $T_{2.k+1}$ with (9) for $P_{k+1}$

3: **if** $L_k = L_{k+1}$ then $T_{0,k+1} = 0$ and $T_{1,k+1} = T - T_{2,k+1}$

4: **else** calculate $T_{0,k+1}$ with (21) and $T_{1,k+1} = T - T_{0,k+1} - T_{2,k+1}$

5: Calculate $bw_i = \frac{s_i}{T_{1,k+1}}$, $i = 1, ..., n$ for $P_{k+1}$ (Theorem 2)

6: Calculate $bw$ for $P_{k+1}$ with (20)

$$bw_{0,k+1} = \frac{2n\lceil \log_2 n \rceil}{T_{0,k+1}} \tag{18}$$

In $T_{1,k+1}$ bandwidth $bw_{1,k+1}$ is used by the base station to send $n$ media blocks to peers, in parallel. From (14) we have:

$$bw_{1,k+1} = \frac{S}{T - T_{0,k+1} - T_{2,k+1}} \tag{19}$$

During $T_{2,k+1}$ the base station bandwidth is $bw_{2,k+1} = 0$.

**Theorem 4**: In the ACIDE dynamic case, $\sum_{i=1}^{n} bw_i$ is minimum if $bw_{0,k+1} = bw_{1,k+1}$. Then, the minimum allocated bandwidth $bw$ is:

$$bw = bw_{0,k+1} = bw_{1,k+1} = \frac{S + 2n\lceil \log_2 n \rceil}{T - T_{2,k+1}} \tag{20}$$

**Proof:** If $bw_{0,k+1} = bw_{1,k+1}$ then $\frac{S}{T - T_{0,k+1} - T_{2,k+1}} = \frac{2n\lceil \log_2 n \rceil}{T_{0,k+1}}$. Time $T_{2.k+1}$ is given by (9). Then, $T_{0,k+1}$ can be calculated as:

$$T_{0,k+1} = \frac{2n\lceil \log_2 n \rceil (T - T_{2,k+1})}{2n\lceil \log_2 n \rceil + S} \tag{21}$$

Because peers receive blocks of optimal sizes $s_i$ in $T_{1,k+1}$, from Theorem 2, $bw_{1,k+1}$ is minimum. From (19) and (21) we have $bw_{1,k+1} = \frac{S + 2n\lceil \log_2 n \rceil}{T - T_{2,k+1}}$. In $T_{0,k+1}$ each peer receives a notification of size $2\lceil log_2 n \rceil$. Then, during $T_{0,k+1} + T_{1,k+1}$ each peer receives a block of optimal size $2\lceil \log_2 n \rceil + s_i$. From Theorem 2, $\sum_{i=1}^{n} bw_i = \frac{S + 2n\lceil \log_2 n \rceil}{T - T_{2,k+1}}$ is the minimum allocated bandwidth during $T_{0,k+1} + T_{1,k+1}$, and $bw = bw_{0,k+1} = bw_{1,k+1}$. □

Algorithm 3 describes the active peer control method. It determines $bw$ for the dynamic case in four steps. In step one $L_{k+1}$ is updated. In step two, the size vector $\mathbf{s}$ for a package $P_{k+1}$ is calculated with (11) and $T_{2.k+1}$ with (9). In step three, $T_{0,k+1}$ and $T_{1,k+1}$ are calculated. In step four, $bw_i$ and $bw$ are determined using Theorem 2 and Theorem 4.

The complexity of Algorithm 3 is given by finding the solution of the system of linear equations in (11) that is $\theta(n^2)$. In Fig. 4 example, $T_{2.k+1}$ is indicated as the upper bound of the *computing time*, the time that it takes Algorithm 3 to complete.

Two observations on extending the *Active Peer Control* method to address bandwidth variations and peers leaving without sending requests are discussed next.

**Observation 4**: Temporal variations in radio propagation conditions, including wireless signal loss, may affect $bw_i$ and $d_i$, $u_i$, $i = 1, ..., n$, throughout Phase 1 and Phase 2 of a very short delay bound *T*. In the static case it is assumed that $bw_i$, $i = 1, ..., n$ remain constant and cluster parameters do not vary during any $T$ of a livestream. In the dynamic case, the APC method performs adaptations to bandwidth variations, and notifications are sent at the beginning of every *T* in response to changes in cluster parameters. For radio perturbations lasting less than one *T*, certain peers may experience reduced $bw_i$ and $d_i$, $u_i$. As a result, a few blocks are lost and it is assumed that the live media quality may experience a slight degradation for a very short period of time. In this scenario, the lost blocks are not retransmitted, and the livestream ratio remains unchanged throughout any delay bound *T*, with a package size *S* maintained constant across all $P_k$, $k = 1, ..., \infty$.

However, temporal variations in radio propagation conditions may persist beyond one *T*, potentially resulting in substantial block loss over multiple consecutive delay bounds. To address these conditions, we propose the *Active Peer Control with Delay Bound Adaptation* (APC-DBA) method. The basic idea is as follows. When bandwidth variations are detected, *T* is divided into a sequence of *m* delay bounds $\tau_j$, $\sum_{j=1}^{m} \tau_j \le T$, such that the cluster parameters *n*, $d_i$, $u_i$ as well as the calculated $bw_i$ remain constant throughout each $\tau_j$. The APC method is used for cluster changes and to send notifications each delay bound $\tau_j$. How bandwidth predictions are performed or how $\tau_j$ are determined is part of future work. A brief overview of the APC-DBA method is given below.

If bandwidth variations are detected during the distribution of a package $P_k$, and blocks are lost, then the subsequent *T* will be divided in several $\tau_j$, $\sum_{j=1}^{m} \tau_j \le T$. Because a package of size *S* has to be distributed in a delay bound $\sum_{j=1}^{m} \tau_j$, then $P_{k+1}$ is divided into a sequence of $m$ packages $P_{k+1,j}$ with sizes $\sigma_j$, $j=1,...,m$, such that $\sum_{j=1}^{m}\sigma_j = S$. A package of size $\sigma_j$ is distributed during $\tau_j$. From Observation 2, the upper bound of the livestream ratio is calculated as $\frac{S}{T} \le \frac{1}{n}\sum_{i=1}^{n} u_i = u_{avg}$. In APC-DBA each $\tau_j$ is determined such that $u_i$, $u_{avg}$ are invariant. Then, the maximum package size $\sigma_j$ that can be distributed during $\tau_j$ is given as follows: $\sigma_j \le \frac{\tau_j}{n}\sum_{i=1}^{n} u_i = \tau_j u_{avg}$.

Furthermore, for each $\tau_j$ the livestream ratio is $\frac{\sigma_j}{\tau_j}$, which implies that the livestream ratio can vary during $T$. To guarantee a continuous media playback however, we should have $\frac{\sum_{j=1}^{m}\sigma_j}{\sum_{j=1}^{m}\tau_j} = \frac{S}{T}$. This means that retransmitting the blocks that were not delivered during a loss of signal time period for example, is possible only if $u_{avg}$ increases in subsequent $\tau_j$. Then, the entire $P_{k+1}$ of size $S$ can be delivered in $\sum_{j=1}^{m}\tau_j \le T$. An example of how $\sigma_j$, $\tau_j$ are adjusting to bandwidth variability is given in section VI.

**Observation 5**: It is possible that during $T$ some peers are leaving a cluster without sending requests and a few blocks are lost. In the worst-case scenario, this could result in the failure to distribute an entire package to peers. The *Active Peer Control with Retransmissions and Bandwidth Reservation* (APC-RBR) is proposed to support livestreaming without live media losses. As shown in Fig. 4, using the *Active Peer Control* method, a package $P_{k+1}$ is distributed in three phases with a delay bound $T$ divided in times $T_{0,k+1}$, $T_{1,k+1}$ and $T_{2,k+1}$. In APC-RBR we propose the retransmission of a package $P_{k+1}$ to peers in an additional Phase 3, during time $T_{3,k+1}$, the *retransmission time*. Then, we have $T = T_{0,k+1} + T_{1,k+1} + T_{2,k+1} + T_{3,k+1}$.

Using time $T_{3,k+1}$ for a package retransmission in Phase 3 reduces the time available for the distribution of $P_{k+1}$ in Phase 1 and Phase 2. As a result, a higher livestream ratio is necessary for a continuous media playback, and the peers are required to sustain higher $bw_i$ $d_i$, $u_i$, and $u_{avg}$ for a shorter period of time in order to receive a package of size $S$. Because $bw$ increases the ACIDE model bandwidth efficiency is reduced.

TABLE II. SIMULATION UPLOAD AND DOWNLOAD BANDWIDTH RANGES

| Cluster Size $n$ | Upload Bandwidth Range [kbps] | $u_{avg}$ [kbps] | Download Bandwidth Range [kbps] |
|---|---|---|---|
| 5 | $U(5)=[10,\ 20]$ | 17.8 | $D(5)=[20,\ 80]$ |
| 10 | $U(10)=[10,\ 30]$ | 22.4 | $D(10)=[30,\ 160]$ |
| 15 | $U(15)=[10,\ 40]$ | 27.3 | $D(15)=[40,\ 240]$ |
| 20 | $U(20)=[10,\ 50]$ | 31.8 | $D(20)=[50,\ 320]$ |
| 40 | $U(40)=[10,\ 60]$ | 44.4 | $D(40)=[60,\ 640]$ |
| 60 | $U(60)=[10,\ 70]$ | 51.7 | $D(60)=[70,\ 960]$ |
| 80 | $U(80)=[10,\ 80]$ | 57.9 | $D(80)=[80,\ 1280]$ |
| 100 | $U(100)=[10,\ 90]$ | 63.5 | $D(100)=[90,\ 1600]$ |
| 120 | $U(120)=[10,\ 100]$ | 68.9 | $D(120)=[100,\ 1920]$ |

The unicast model or the ACIDE model can be used to retransmit $P_{k+1}$ to the remaining peers. Even if the number of peers is reduced, when the unicast model is used, a significant large bandwidth is required. If there is not enough bandwidth available during $T_{3,k+1}$, a retransmission is not possible. A more efficient alternative is to reserve bandwidth during $T_{3,k+1}$ and use Algorithm 2 for the number of peers optimization. A simulation example is presented in section VI.

## VI. SIMULATION AND PERFORMANCE EVALUATION

In this section the performance evaluation of the ACIDE media distribution model is presented. The setup description is followed by discussions on Problem 1 simulation for the static and dynamic cases. The mesh and star configuration results are identical. Problem 2 performance is analyzed in [6].

### A. Simulation Methods and Setup Description

GNU Octave, a set of tools designed for solving linear algebra problems, has been used for simulation.

A delay bound of $T$ = 200ms and cluster sizes of $n \in \{5,10,15,20,40,60,80,100,120\}$ peers have been chosen for the bandwidth optimization simulation. For each $n$, the upload and download bandwidth ranges $U(n)$ and $D(n)$ respectively, are listed in Table II. For $n$ = 5 for example, the notation $U(5)=[10,20]$kbps means that $10 \le u_i \le 20$ kbps. The $U(n)$ and $D(n)$ ranges have been selected to emphasize two scenarios: 1) the increase of mobile devices $u_i$ and $d_i$ made possible by technological advancement and 2) the $u_i$ and $d_i$ changes due to varying radio propagation conditions. From the above, the $U(n)$ ranges have been defined to satisfy the following: $U(5) \subset ... \subset U(120)$ and $U(n) \cap U(n+1) = U(n)$. The upper limits of $D(n)$ are equal to $n$ times the largest $\frac{S}{T}$.

Parameters $u_i$ and $d_i$ have been chosen at random, from Table II ranges $U(n)$ and $D(n)$ respectively. It is assumed that $u_1 \le ... \le u_n \le \min\{d_1,...,d_n\}$. The $u_{avg}$ values of all $U(n)$ are presented in Table II. To find how changes in the quality of livestream media influence the minimum allocated bandwidth results, the livestream ratios $\frac{S}{T} \in \{10,12,14,16\}$ kbps have been selected such that Observation 2 is true for $n$ = 5 and the $u_{avg}$ of $U(5)$. Therefore, $\max\{10,12,14,16\}$ kbps $\le 17.8$ kbps. Livestream ratios in the $[10, 20]$ kbps range are similar to those currently supported by some popular platforms livestreaming social media content, radio and TV channels. The conclusion remains unchanged even when significantly larger bandwidth values are used.

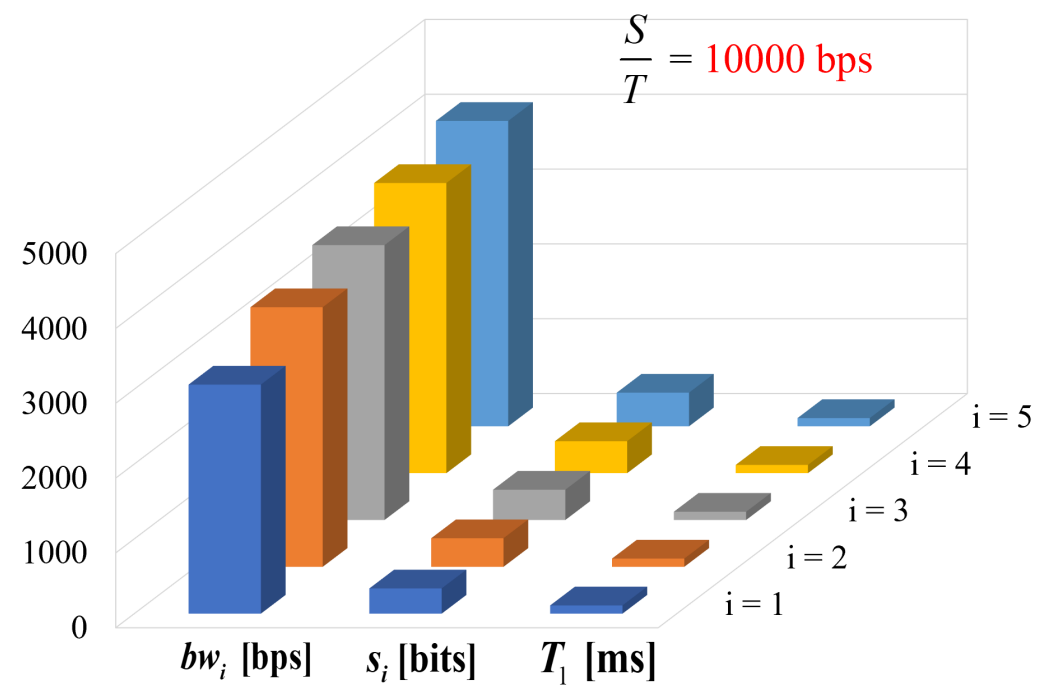

Fig. 5. Block sizes and bandwidth allocated to $n=5$ peers

For the given livestream ratios and Table II parameters, the package sizes defining the quality of a livestream media are $S \in \{2000, 2400, 2800, 3200\}$ bits, with $S \le 3560$ bits.

### B. Bandwidth Optimization Problem Simulation Results

The purpose of this simulation is evaluating the variation of the allocated bandwidth $bw$ as $n$ is getting larger and $u_{avg}$ increases. Algorithm 1 has been used to find the optimal block sizes $\mathbf{s} = [s_1,...,s_n]^t$. This solution is used to calculate $bw$, $T_1$, $T_2$, and the optimal values of $\mathbf{bw} = [bw_1,...,bw_n]^t$.

According to Theorem 1, the events in Phase 1 take the same time. The result is displayed in Fig. 5 for a multi-channel radio, where the block size to bandwidth ratios $\frac{s_i}{bw_i}$ are equal to $T_1$ for all $n$ = 5 peers. For the example presented in Fig. 5, $\mathbf{bw} = [3.061, 3.469, 3.673, 3.877, 4.081]^t$ kbps values have been allocated to peers $i = 1,...,5$ to download blocks of sizes $\mathbf{s} = [337, 382, 404, 426, 449]^t$ bits, in parallel, in $T_1 = 111$ ms.

Fig. 6 shows the allocated bandwidth variation with $n$ and it points out that as $n$ and $u_{avg}$ are getting larger, $bw$ is decreasing and the base station bandwidth is utilized more efficiently. *The first important result* is that for a large $n$ and increasing $u_{avg}$ the values of $bw$ are getting closer to $\frac{S}{T}$. Fig. 6 points out that for a fixed $n$ the ACIDE model efficiency decreases as the livestream ratio approaches its upper bound, $u_{avg}$. When the livestream ratio reaches $\frac{S}{T} = u_{avg}$ then $bw = n\frac{S}{T}$, the unicast bandwidth, and the ACIDE model becomes ineffective.

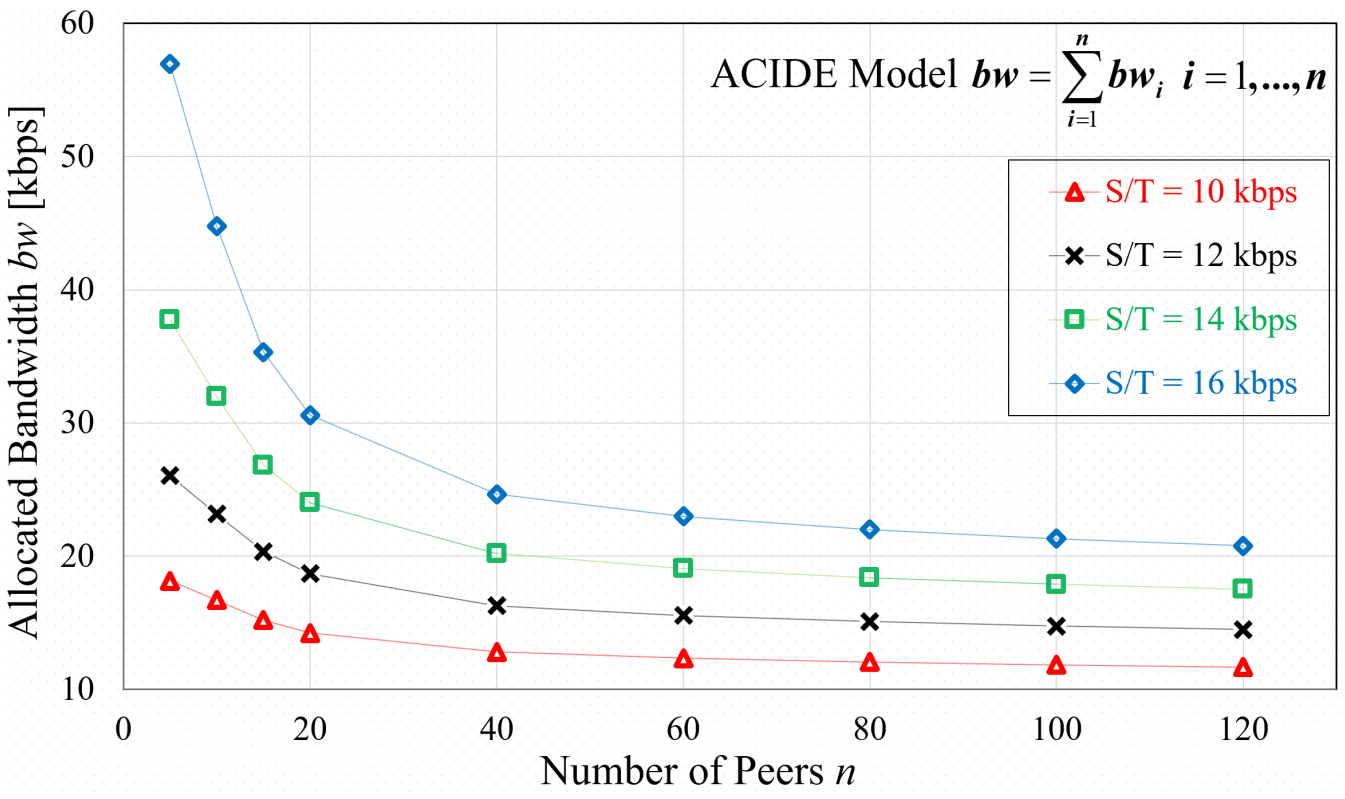

Fig. 6. Allocated bandwidth $bw$ variation with $n$, $u_{avg}$ and $\frac{S}{T}$

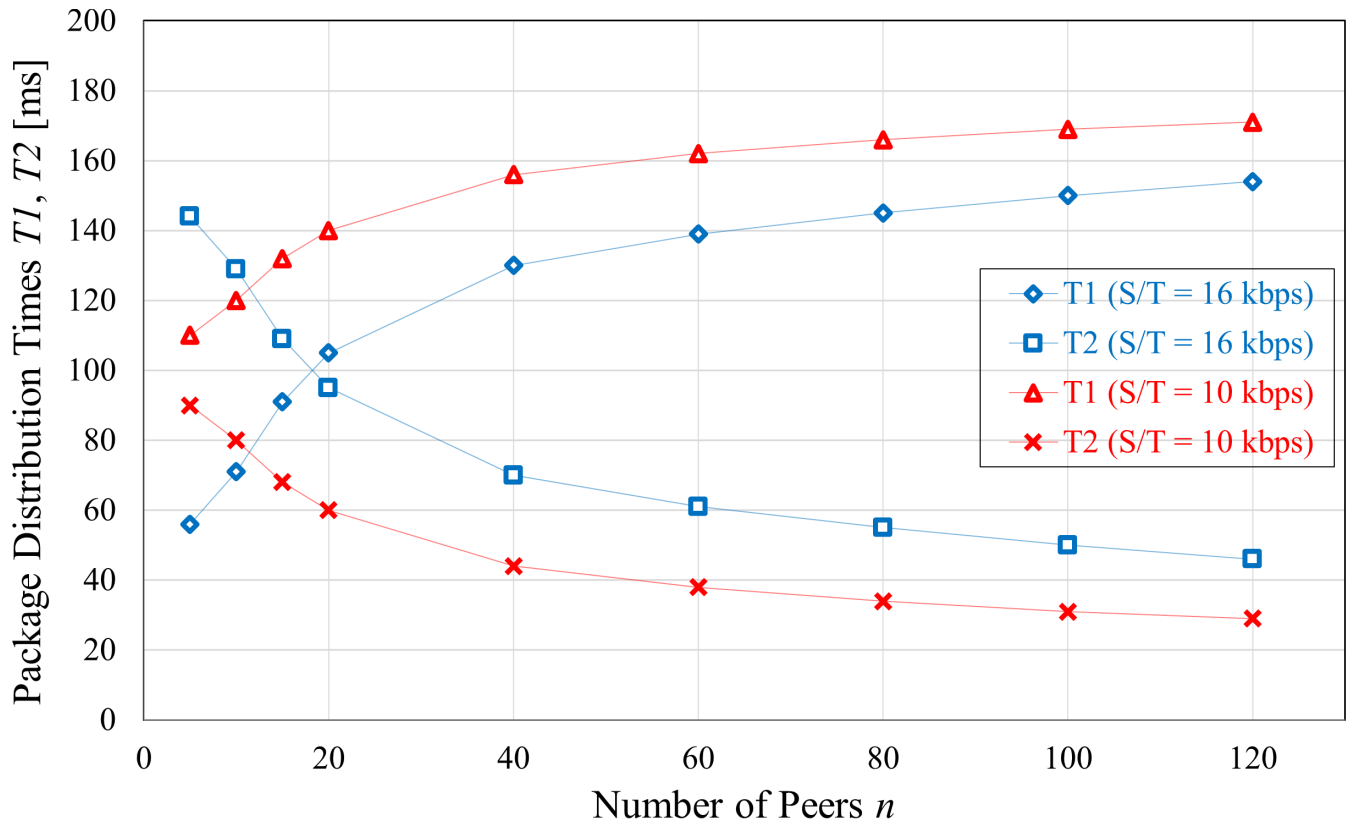

Fig. 7. Package distribution times $T_1$, $T_2$ variation with $n$, $u_{avg}$ and $\frac{S}{T}$

Fig. 7 indicates that as the livestream ratio is getting larger $T_1$ is decreasing, meaning $bw$ is increasing with the livestream ratio $\frac{S}{T}$. Then, according to Assumption 3, $bw \ge \frac{S}{T_1} > \frac{S}{T}$. For example, in Fig. 6 and Fig. 7, for $n$ = 60 we have $bw$ = 12.34 kbps and $T_1$=162ms, $T_2$=38ms if $\frac{S}{T}$=10kbps. For a livestream ratio of $\frac{S}{T}$=16kbps, $bw$ = 23 kbps and $T_1$=139ms, $T_2$=61ms.

Moreover, for a fixed livestream ratio, $T_2$ is reduced as $n$, $u_{avg}$ increase, implying that in order to guarantee a constant $T$, time $T_1$ should increase. From Theorem 1 and (5) this process is inducing the reduction of $bw$. For example, in Fig. 6 and Fig. 7 for a 10kbps livestream ratio, $bw$=12.34 kbps, $T_1$=162ms, $T_2$=38ms for $n$=60 and $bw$=11.69 kbps, $T_1$=171ms, $T_2$=29ms for $n$=120. Our *second important result*, is that $T_2$ is the allocated bandwidth control loop variable. Consequently, the set of peers admitted to a cluster can directly control $bw$.

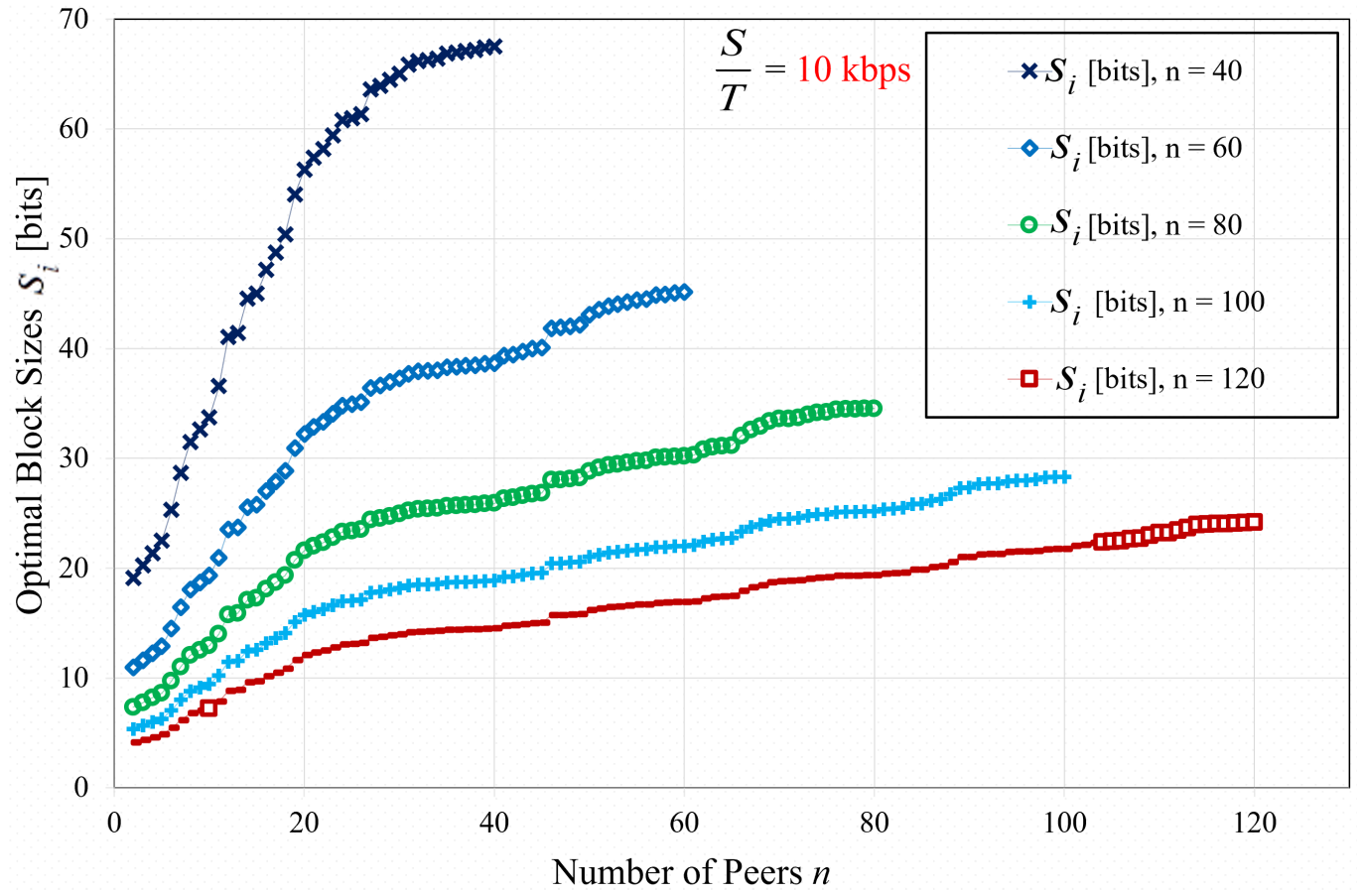

Fig. 8. Optimal block sizes $s_i$ changes with $n$ and $u_{avg}$ for $\frac{S}{T} = 10\text{kbps}$

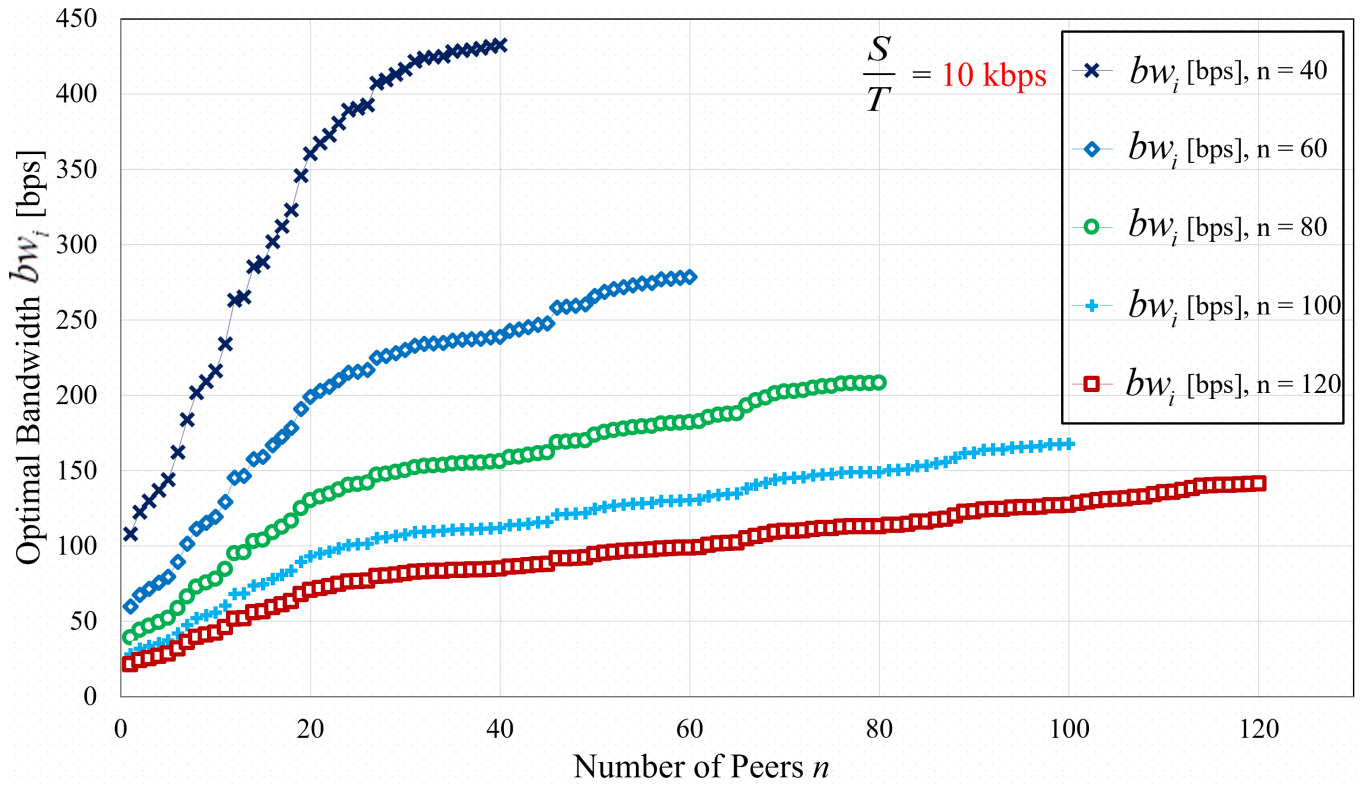

Fig. 9. Minimum bandwidth $bw_i$ changes with $n$ and $u_{avg}$ for $\frac{S}{T} = 10\text{kbps}$

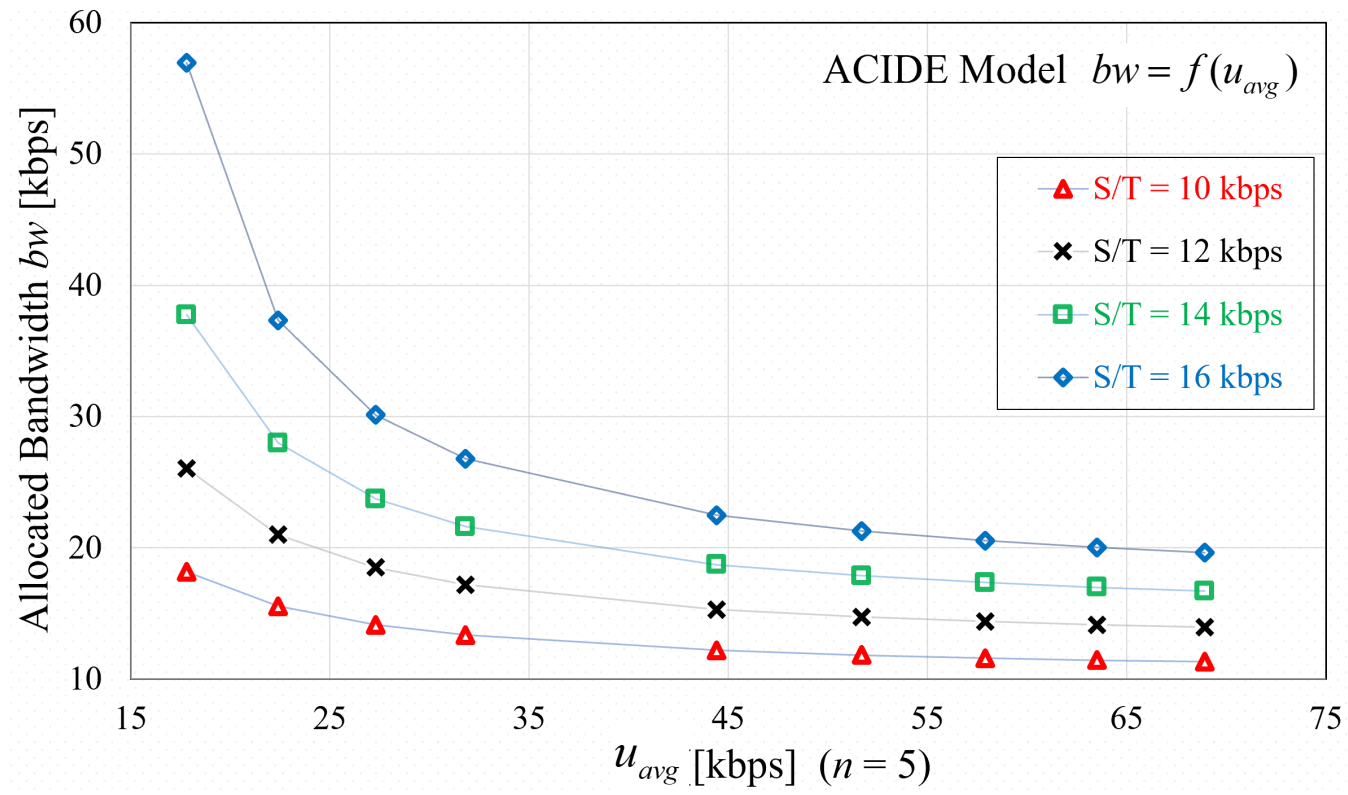

Fig. 10. Allocated bandwidth $bw$ variation with $u_{avg}$, $\frac{S}{T}$ for a constant $n$ = 5

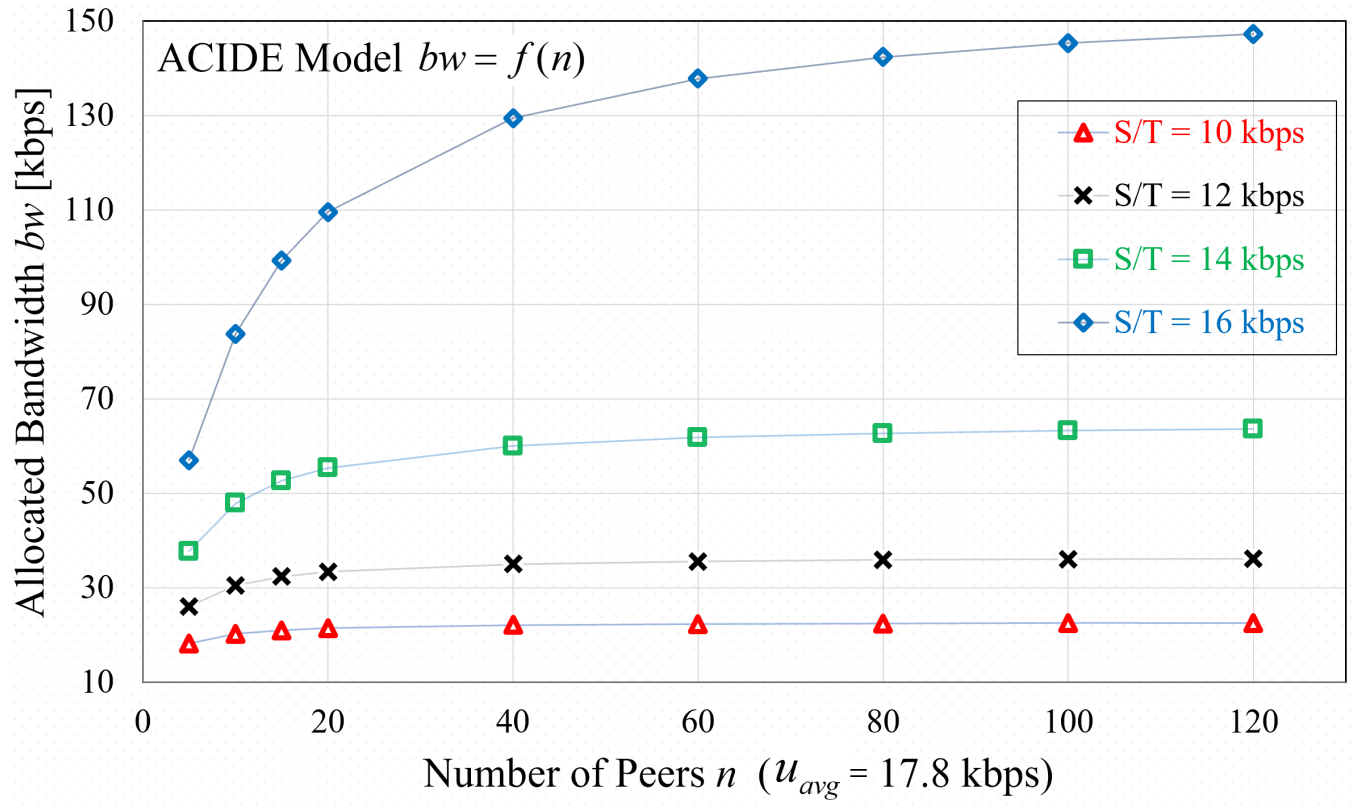

Fig. 11. Allocated bandwidth $bw$ variation with $n$, $\frac{S}{T}$ for $u_{avg} = 17.8$ kbps

Fig. 8 and Fig. 9 point out that the average rates of change for $s_i$ and $bw_i$ respectively are decreasing as $n$ and $u_{avg}$ are getting larger. Fig. 9 demonstrates that the minimum bandwidth allocated to send a block with the optimal size $s_i$ to a peer $i$ is less than peer's $i$ download bandwidth, that is $bw_i \le d_i$. For $n$ = 40 for example, we have $bw_i \le \min\{d_1,...,d_{40}\} = 60$ kbps. The *third important result* is that $s_i$ and $bw_i$, $i = 1,...,n$, decrease as $n$ and $u_{avg}$ are getting larger [6].

A simultaneous increase of $n$ and the corresponding $u_{avg}$ has been assumed for the results presented in Fig. 6 through Fig. 9. For the results in Fig. 10 and Fig. 11, $n$ and $u_{avg}$ are varied independently. Fig. 10 demonstrates the second result from (15) in Observation 1, that is, $bw$ is approaching $\frac{S}{T}$ if $u_{avg}$ is getting larger and $n$ is constant. The third result of Observation 1 is illustrated in Fig. 11, where for a large number of peers and a fixed $u_{avg}$, $bw$ is approaching the value in (16) $\dfrac{S}{T - \dfrac{S}{u_{avg}}} > \dfrac{S}{T}$.

Fig. 10 indicates a minimum allocated bandwidth $bw$ =19.651kbps for $\frac{S}{T}$ = 16kbps, $n$=5 and $u_{avg}$ = 68.9kbps. The result is less than the 80kbps, the bandwidth necessary if a unicast model would be used for livestreaming when $\frac{S}{T}$ = 16kbps, $n$=5. In Fig. 11, if $\frac{S}{T}$ = 16kbps, $n$ =120, and $u_{avg}$ = 17.8kbps, we have $bw$ = 147.31kbps compared to a unicast value of 1920kbps.

We observe that a more significant $bw$ reduction from the unicast value is achieved for a larger $n$. The *first important result* of our simulation points toward an improved livestreaming bandwidth efficiency when $n$ and $u_{avg}$ are getting larger at the same time. In Fig. 6 for example $bw$ = 20.782kbps for $\frac{S}{T}$ = 16kbps, $n$ =120, and $u_{avg}$ = 68.9kbps.

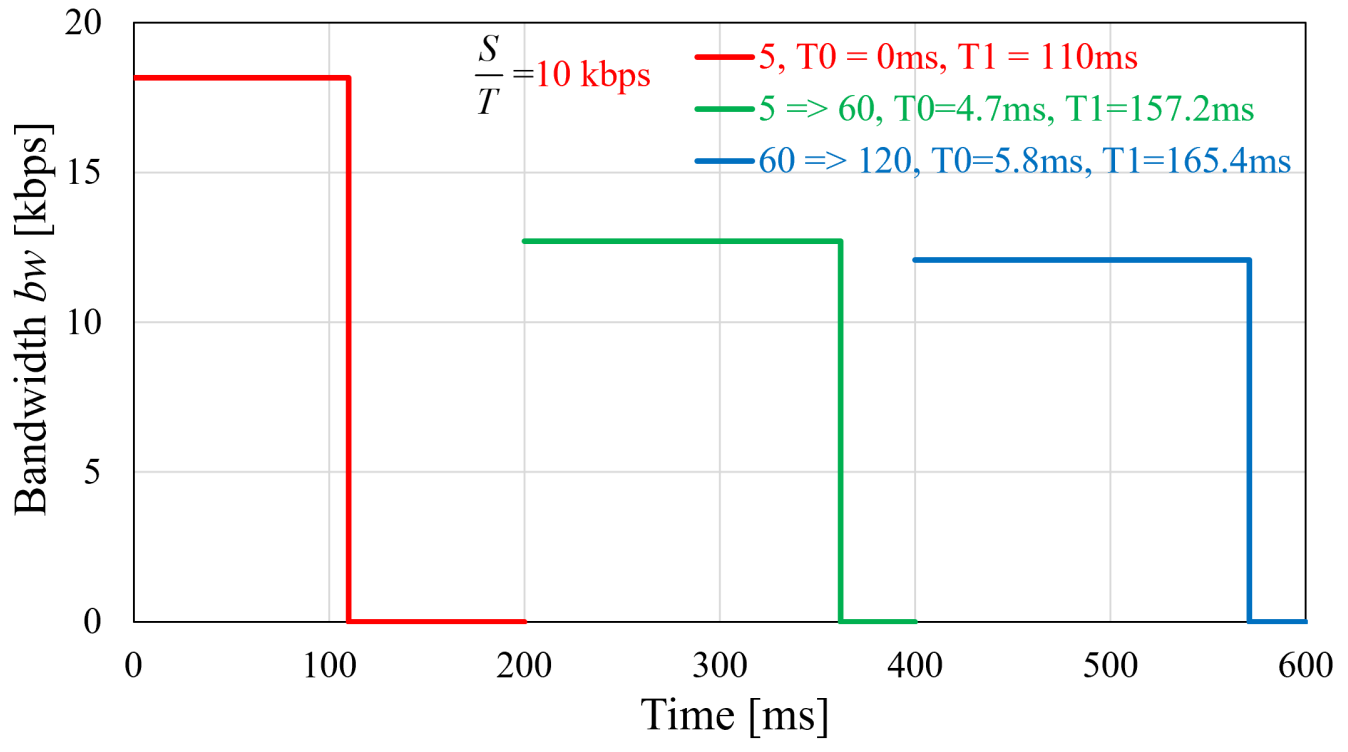

Fig. 12. Allocated bandwidth $bw$ in the dynamic case ($n$ and $u_{avg}$ increase)

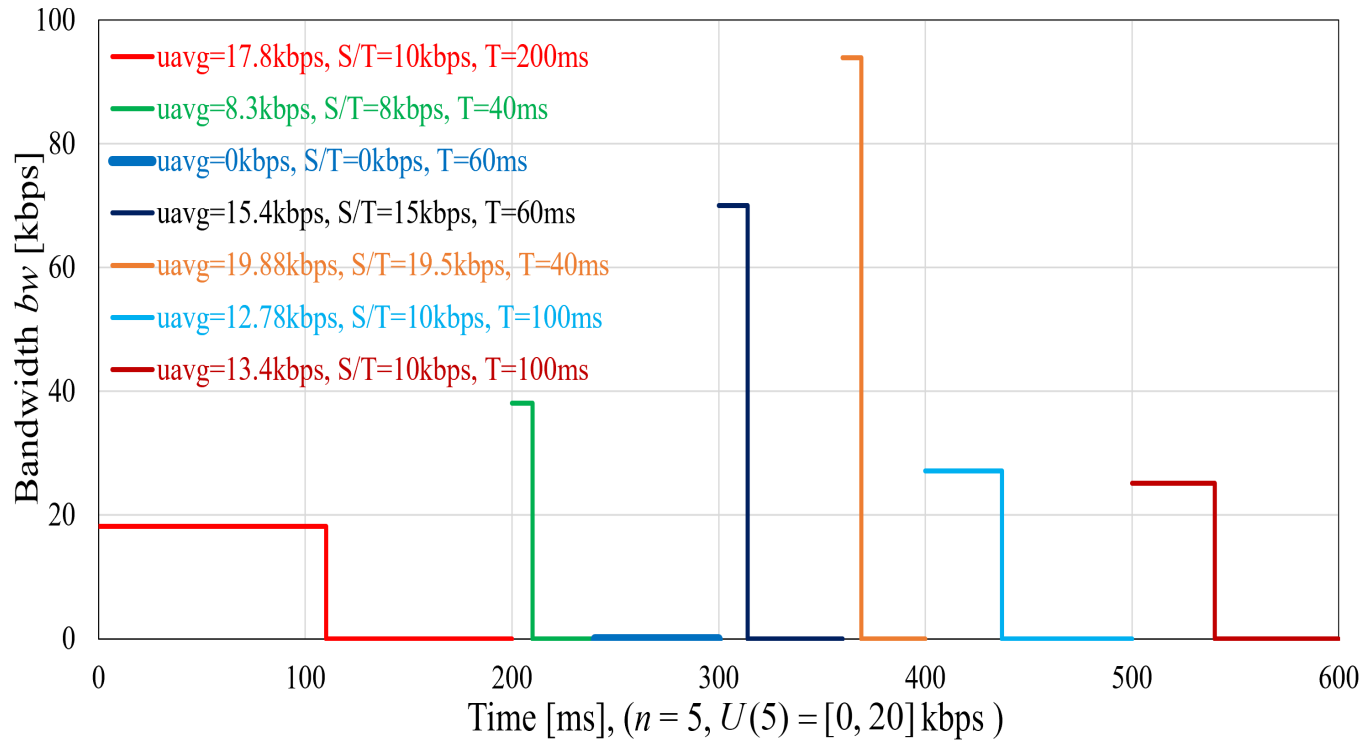

Fig. 13. An example of delay bound adaptation to bandwidth variations

### C. *Active Peer Control Bandwidth Optimization Simulation*

In this section the dynamic case simulation results are presented. Algorithm 3 has been used to calculate $bw$ when new peers are joining a cluster having an initial size $n$ = 5 peers. The media livestream is divided in packages of size $S$ = 2000bits. This cluster is distributing $P_1$ in $T$ = 200ms. It is assumed that $T_{0,1}$=0ms and $\frac{S}{T}$ =10kbps. In two consecutive delay bounds $T$, 55 and 60 new peers are joining the cluster. Then, packages $P_2$ and $P_3$ are distributed to $n$ = 60 and $n$ = 120 peers respectively.

The results in Fig. 6 for $n$ = 5, $n$ = 60 and $n$ = 120, with the corresponding ranges and $u_{avg}$ given in Table II, have been used as a baseline to demonstrate how the $bw$ of a changing cluster increase in the dynamic case. A cluster change from $n$ = 5 to $n$ = 60 for example, implies that 55 new peers with $20 \le u_i \le 70$ kbps are joining the existing five peers. After the change, the cluster has 60 peers with ranges $U(60)$, $D(60)$, and an $u_{avg} = 51.7$ kbps.

Fig. 12 points out that $bw = bw_{0,2} = bw_{1,2} = 12.71$ kbps for $P_2$ and $bw = bw_{0,3} = bw_{1,3} = 12.08$ kbps for $P_3$. Because more peers are added to the cluster and $u_{avg}$ is getting larger, the times for Phase 2 are $T_{2,1} > T_{2,2} > T_{2,3}$ and $bw_{1,1} > bw_{1,2} > bw_{1,3}$. For $n$ = 60, since $T_{0,2} \neq 0$ we notice a $bw$ increase from 12.34 kbps in the static case to 12.71 kbps in the dynamic case. Similarly, for $n$ = 120, $bw$ is increasing from 11.69 kbps in the static case to 12.08 kbps in the dynamic case.

For both static and dynamic cases $T_{2,k}$ is identical because, for a fixed $\frac{S}{T}$, $T_{2,k}$ depends only on $n$ and $u_{avg}$. When the notification time $T_{0,k} \neq 0$ in the dynamic case, the blocks of a package $P_k$ are distributed in $T_{1,k}$, a time shorter than the time used in the static case. As a result, $bw$ is larger. For example, for $n$ = 60, $T_{2,2} = 38.1$ ms in both static and dynamic cases. Because $T_{1,2} = 157.2$ ms, which 4.7ms less than Phase 1 time in the static case, $bw$ is increasing with 370bps.

Another observation from Fig. 12 is as follows. It may also be assumed that 115 new peers join the initial cluster of $n$ = 5. Algorithm 3 results in this case are identical to the previous case where 60 new peers are joining the $n$ = 60 cluster because the base station is sending notifications to all $n$ = 120 peers in time $T_{0,3}$. As indicated in Fig. 12, because $T_{2,1} > T_{2,2}$, the upper bound of the base station computing time is lower for the case where 60 new peers join the cluster with $n$ = 60. Therefore, in high user density mobile wireless networks it is more bandwidth efficient to add a large number of peers to smaller size clusters.

Fig. 13 illustrates an example of delay bound adaptation to bandwidth variability. In this simulation, $\sigma_j$ and $\tau_j$, as defined in Observation 4, serve as input variables to Algorithm 3. The following scenario is considered. While $P_1$ is distributed, the radio propagation conditions for $n$ = 5 peers vary during the time interval 0 < $t$ < 200ms, and $u_{avg}$ is reduced to 8.3kbps. Subsequently, $P_2$ is divided into a sequence of packages $P_{2,j}$, $j = 1,...,m$. Fig. 13 illustrates the distribution of packages $P_{2,1}$, $P_{2,2}$, $P_{2,3}$, and $P_{2,4}$ during the time interval 200ms < $t$ < 400ms. For $P_{2,1}$, parameters $d_i$, $u_i$, $\sigma_1$, $\tau_1$ are updated, where $d_i$, $u_i$ denote the maximum sustainable bandwidth during $\tau_1$. At $t$ = 200ms, $P_{2,1}$ time is $\tau_1$ = 40ms, the livestream ratio is 8kbps and $\sigma_1$ = 320bits. Then, a loss of wireless signal is detected during the distribution of $P_{2,1}$. Thereafter, the updated parameters for $P_{2,2}$ are $d_i = u_i$ =0kbps, $\sigma_2$ = 0bits and $\tau_2$ = 60ms. During 240 < $t$ < 300ms the propagation conditions are restored and parameters are updated. At $t$ = 300ms, the distribution of $P_{2,3}$ is initiated, followed by that of $P_{2,4}$. During $\tau_3$ = 60ms and $\tau_4$ = 40ms a total of 1680bits are received by peers. Then, the entire $P_2$ is distributed within $T$, and

$$\frac{\sum_{j=1}^{4} \sigma_j}{\sum_{j=1}^{4} \tau_j} = \frac{320+0+900+780}{40+60+60+40} = \frac{S}{T}$$

. It is assumed that an increase in $u_{avg}$ results in livestream ratios of 15kbps for $P_{2,3}$ and 19.5kbps for $P_{2,4}$, guaranteeing a continuous playback without a change in quality. During the time interval 400ms < $t$ < 600ms, package $P_3$ is divided in $P_{3,1}$ and $P_{3,2}$ for distribution in $\tau_1$ and $\tau_2$, illustrating a strategy for adapting the delay bound in response to bandwidth variations across successive *T*. Since $d_i$, $u_i$ cannot be sustained longer than 100ms, then $\tau_1 = \tau_2 =$ 100ms and $P_{3,1}$, $P_{3,2}$ are distributing $S$ = 2000bits in $T$ = 200ms.

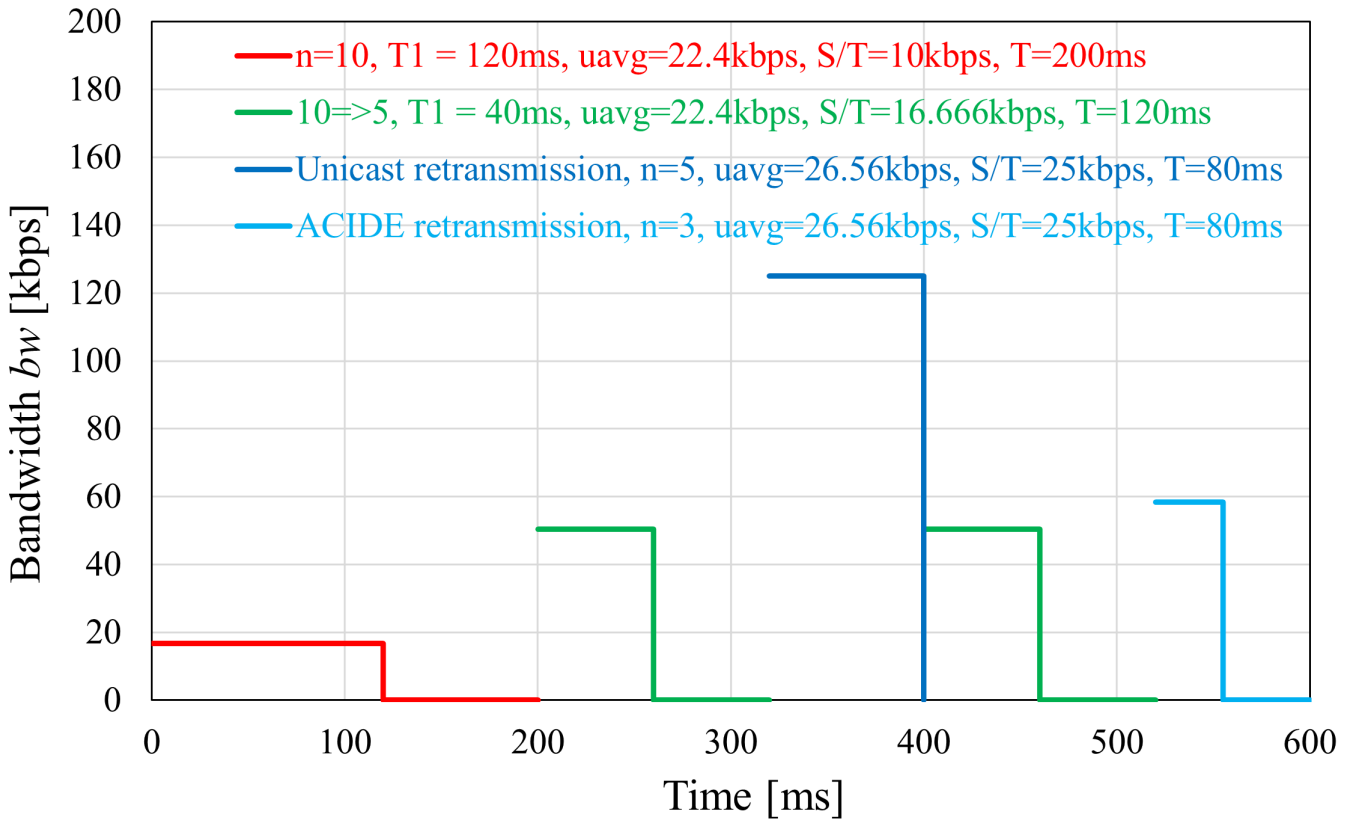

Fig. 14. Active Peer Control with Retransmissions and Bandwidth Reservation

One notable observation from Fig. 13 is that, following a loss of signal, the livestream ratio is constrained by $u_{avg}$. Consequently, the retransmission of the lost package $P_{2,2}$ may become infeasible and the live media quality may be degraded if $\sigma_3 + \sigma_4 > \tau_3 u_{avg,3} + \tau_4 u_{avg,4}$.

The simulation results of a package distribution while using the APC-RBR method are shown in Fig. 14. During 0ms < $t$ < 200ms, $n$ = 10 peers are livestreaming without using RBR. It is assumed that during both 200ms < $t$ < 400ms and 400ms < $t$ < 600ms time intervals, RBR is used and five peers are leaving the cluster during the distribution of a package, resulting in lost media blocks. The two package retransmission options during $T_3$, corresponding to the unicast and the ACIDE models, are presented for comparison analysis across time intervals 200ms < $t$ < 400ms and 400ms < $t$ < 600ms respectively.

To allocate $T_3$ for retransmissions, it is assumed that the distribution of a package with RBR is completed within a shorter delay bound, that is $T_1$=120ms, the time of Phase 1 from 0ms < $t$ < 200ms interval. Therefore, in the RBR case, *bw* increases from 16.716kbps to 50.444kbps. In the scenario shown in Fig. 14, $T_3$ =80ms and a livestream bandwidth of 25kbps is used for the distribution of a 2000bits package. If no retransmissions are required, $T_3$ can be used for other communications. The unicast model requires the allocation of a large amount of bandwidth. In many cases, it is more efficient to reserve a portion of the available bandwidth for retransmissions during $T_3$. The number of peers admitted to a cluster can be optimized using Algorithm 2. For the example in Fig. 14, a bandwidth of 60kbps is reserved. However, according to the ACIDE model, only $n$ = 3 peers can be admitted to the cluster with a *bw* of 58.339kbps. Furthermore, the delay bound adaptation method can be integrated with the RBR strategy to enhance the overall livestreaming reliability.

## VII. Related Work

Our work aims to provide a computationally feasible framework supporting the self-controlled optimization of livestreaming bandwidth and number of users in mobile wireless networks. For a continuous play, the live media should be distributed to users within a delay bound. This is difficult to achieve in mobile wireless networks where large number of users and varying conditions on radio propagation increase the variability of wireless bandwidth. To address these challenges, we have analyzed different approaches using P2P communications, multicast and content delivery techniques.

Much research work on P2P networks has been dedicated to the design of bandwidth efficient overlay routing architectures optimizing media streaming to users [8], [9], [10]. Studies on P2P mobile wireless networks indicate that the available bandwidth per user is reduced when the number of users is increasing [11], [12]. One solution proposed for an efficient network bandwidth utilization is to provide mostly short-range communications [11]. Another approach is to allow a large number of peers and use a non-linear, utility-based model for bandwidth allocation to peers [13], [14].

Multicast research studies in mobile wireless networks propose distributed, content coded caching techniques and cluster grouping schemes [15], [16], [17], [18] [19]. Media is divided, coded and cached on many network devices before being distributed to the users. Based on their content interest, users can be grouped in clusters for improved delay bound media distribution. Heterogeneous wireless networks are evaluated for bandwidth efficient livestreaming in [20]. For a large number of users, these networks may implement the solution described in [11] because users have the ability to configure their interfaces as multi technology [21] or multi-protocol [22] and connect to many networks or cluster together [23], [24], [25]. Hybrid content delivery networks and P2P solutions are analyzed in [26], [27]. Recorded media is divided in chunks and distributed to network devices. Peers are using caching techniques to download a subset of chunks from the cached content and the remaining chunks from a P2P cluster.

The problem of bandwidth efficient livestreaming in high user density mobile wireless networks has not been widely researched and no solutions are available in the literature. Very few of the existing studies attempt solving the bandwidth and capacity optimization problems during heavy livestreaming demand. For example, the bandwidth utilization and its variability have been identified as important challenges and overlay routing solutions have been proposed in [28], [29] to improve transmission performance between nodes. Non-linear algorithms have been analyzed in [30], [31], [32] for nodes resource optimization in bandwidth variable P2P communications, and node selection algorithms have been proposed for increasing P2P collaboration efficiency [33], [34]. Data distribution performance studies on P2P wireless and wired overlay topologies indicate that the energy consumption, bandwidth utilization and latency increase with the number of participating nodes [35], [36]. Resource allocation, user grouping algorithms and media multicast transmission methods in wireless networks have been presented in [37], [38], [39] in an attempt to reduce the bandwidth utilized by video streaming applications. Better performance results on energy consumption and bandwidth utilization for high user density wireless networks have been reached using D2D solutions, or with multicast user grouping algorithms in hybrid and wireless heterogeneous networks [40], [41], [42]. Other approaches propose to dynamically adapt the source livestream ratio to the time varying throughput of mobile edge networks [43], [44] or study techniques of latency reduction and packets delivery under time constraints using base station clustering and resource allocation schemes [45], [46], [47], [48].

The closest related work is the media distribution model proposed in [4], [5], in which a server is distributing recorded media over a wireline network to P2P users. The problem of dividing a media object in segments is addressed. The model calculates the segments optimal sizes such that their distribution time to peers is minimized. It is assumed that peers use their upper bound download bandwidth when exchanging segments over $\frac{n(n-1)}{2}$ concurrent bidirectional connections. An optimal solution of a linear program is found by the simplex method.

The model proposed in this study borrows some concepts from the media object segmentation of [4], [5]. The ACIDE model is used for media livestreaming in mobile wireless networks, a different application running in highly variable bandwidth conditions. Using base station resources more efficiently and increasing the network capacity as long as live media is played on mobile devices with no interruptions are the objectives of the model. Our approach does not use an overlay and all interested peers are collaborating on media distribution. The basic idea is to group $n$ users located in the proximity of each other and interested in the same media, in a cluster of peers. Inside a cluster, $n$ peers are able to establish short-range P2P communications, using a frequency range outside the base station frequency band. A livestream is sent in packages. Each package is divided into $n$ blocks of optimal sizes. The blocks are delivered to $n$ peers within a constant delay bound, in two-phases. In Phase 1, each peer receives one block from the base station and in Phase 2 the peers exchange their blocks and reconstruct the package. As the number of peers and $u_{avg}$ increase, the bandwidth allocated to each peer in Phase 1 is reduced, making it less susceptible to radio propagation variations. Additionally, to reduce the interference between P2P communications, each peer can utilize no more than two unidirectional connections throughout Phase 2 for the transfer of a total of $n$-1 blocks. A comparison with the most closely related papers is given in Appendix C. Further discussions on the ACIDE P2P communications follow.

The challenges associated with multimedia broadcast and multicast services in both existing and emerging mobile networks are examined in [49], [50]. Research has primarily focused on improving data rates, extending the coverage, energy efficiency and reducing latency. There has been limited attention given to the capacity optimization problem, particularly in high density environments.

The first major challenge in implementing broadcast or multicast over broadcast is the variation of radio channel conditions among users accessing the same content. As a result, network capacity is diminished, as users with a $d_i$ lower than the livestream ratio are unable to receive the live media. Proposed solutions such as multicast with network coding and link adaptation schemes have been suggested to address this issue. Collaborative D2D techniques, where users with $d_i$ higher than the livestream ratio act as relays between the base station and users with lower $d_i$ have been also presented. However, these solutions are not recommended for livestreaming due to time constraints. The complexity of network coding and relays increases latency, making it difficult to guarantee a continuous live media playback. Other approaches, such as multi-antenna MIMO and beamforming, can deliver high bandwidth and support a large number of users [51], [52]. However, they are highly complex and demand a significant infrastructure investment. Collaborative strategies, short-range communications, and joint unicast-multicast methods are identified as potentially more efficient alternatives.

The second major challenge is the inefficient use of radio frequency bands when channels are dedicated solely for livestreaming or vehicular communications for example. Although livestreaming a TV channel may be bandwidth-efficient in this case, the network capacity remains limited, as only users interested in the same content are sharing the communication channel. Moreover, social media livestreams are typically short lived, further compounding the challenge. In practical applications, communication channels must accommodate broadcast, multicast, and unicast traffic, meaning that bandwidth must be allocated to each type of traffic accordingly. Let's assume that for the example given in Section I $B$, one channel is shared and $T$ should be divided between a cluster with $n$ = 10 robotaxis, $\frac{S}{T}$ = 16kbps, a cluster with $n$ = 40 peers at the festival livestreaming the same content with $\frac{S}{T}$ = 10kbps, and unicast users. For a $T$ = 200ms and the Table II bandwidth values, then $bw$ = 44.8kbps is allocated to the robotaxis for $T_1 = 71$ms and $bw$ = 12.8kbps is allocated to the 40 peers for $T_1 = 156$ ms. Clearly, from Observation 3 and (17) $bw$ needs to increase in order to reduce the channel access time. For instance, if $bw$ doubles for both clusters, then $T_1 = 35.5$ ms for the robotaxis and $T_1 = 78$ms for the cluster of $n$ = 40 peers, leaving 86.5ms to the unicast users. The key observation is that for shared channels the ACIDE bandwidth allocation equals the multicast bandwidth. This fundamental result holds true weather blocks are sent using serial or parallel communication modes on shared single-channel or multi-channel radios.

The two-phase unicast-multipoint ACIDE communication model suggests a more bandwidth efficient livestreaming in high density networks by actively controlling the number of users. By dividing a package in optimal blocks according to $d_i$, $u_i$, and using P2P communications in Phase 2, peers with lower $d_i$ can take advantage of the excess bandwidth available to peers with more resources. This enables the dynamic control of bandwidth efficiency by adjusting the number of peers and the average upload bandwidth. Furthermore, live media quality and reliability may be improved by continuously adapting the size and delay bound of a package, and by integrating package retransmissions with bandwidth reservation strategies.

The objective of this work is to establish the theoretical results of the ACIDE model, therefore, detailed P2P cluster channel modeling or security mechanisms are beyond the scope of this paper. Appendix D briefly discusses a potential physical-layer realization to assess the feasibility and efficiency of the proposed ACIDE model.

## VIII. Conclusion

In this study, the ACIDE model is proposed to improve the livestreaming bandwidth efficiency in mobile wireless networks. The model aims to minimize the base station bandwidth needed to guarantee an uninterrupted live media play. We formulated the bandwidth minimization problem and identified the optimal conditions for dividing and distributing a livestream package as *n* media blocks to the *n* peers of a cluster. Our proposed solution has low complexity and is able to find the optimal media block sizes by solving a system of linear equations. Additionally, the allocated bandwidth is reduced as the size of a cluster and its average upload bandwidth are getting larger. A greedy strategy is proposed for solving the NP-complete network capacity optimization problem. For a known reserved bandwidth, the model is able to calculate a feasible solution for peer selection. The *Active Peer Control* method has been proposed to dynamically update a cluster when peers join or leave.

The ACIDE model improves the wireless bandwidth efficiency. As a result, more users are allowed and the network capacity increases. The effects of bandwidth variability are being further analyzed.

## Appendix A

### Proof of Theorem 1

**1.** If **bw**, **s** are optimal then all events in Phase 1 have equal times and in Phase 2 for each step all events have equal times.

*Phase 1*: First, we prove that if $\mathbf{bw} = [bw_1,...,bw_n]^t$ and $\mathbf{s} = [s_1,...,s_n]^t$ are optimal then $T_1 = \frac{s_1}{bw_1}.=..=\frac{s_n}{bw_n}$. We assume that Phase 1 block distribution times are not equal.

Without loss of generality, let the following time ratios $\frac{s_1}{bw_1} = ... = \frac{s_l}{bw_l} = \max\{\frac{s_i}{bw_i}, i = 1,...,n\}$ (in other words we have $l$ maximum ratios) and $\frac{s_n}{bw_n} = \min\{\frac{s_i}{bw_i}, i = 1,...,n\}$. Since the bandwidth distribution in Phase 1 does not affect $T_2$, we can redistribute the bandwidth to shorten the maximum ratios (time).

Let $bw_i^{'} = bw_i + \delta_i$, for $i = 1,...,l$, such that we have $\frac{s_n}{bw_n - \sum_{i=1}^{l}\delta_i} \le \frac{s_1}{bw_1 + \delta_1} = ... = \frac{s_l}{bw_l + \delta_l}$. Then there is another solution with the same amount of bandwidth $\sum_{i=1}^{n} bw_i$, and a shorter time. This implies that the given solution is not optimal. Therefore, if **bw** and **s** are optimal the events in Phase 1 have equal times.

*Phase 2*: We assume that the event completion times in step $k$, $k = 1,...,n-1$, are not equal. Let the following time ratios $\frac{s_1}{u_1} = ... = \frac{s_m}{u_m} = \max\{\frac{s_i}{u_i}, i = 1,...,n\}$, so we have $m$ equal, maximum times, and $\frac{s_n}{u_n} = \min\{\frac{s_i}{u_i}, i = 1,...,n\}$. We can change the block sizes allocated to each peer to reduce the time that events in step $k$ take to finish. Let $s_i^{'} = s_i - \gamma_i$, $i = 1,...,m$, such that $\frac{s_n + \sum_{i=1}^{m}\gamma_i}{u_n} \le \frac{s_1 - \gamma_1}{u_1} = ... = \frac{s_m - \gamma_m}{u_m}$. Then, there is another solution $\mathbf{s}^{'}$ with the same size $\sum_{i=1}^{n} s_i$ and a shorter time. This proves that if **bw** and **s** are optimal the events in step $k$ of Phase 2 have equal times. The result is valid for all *n-1* steps.

**2.** If all events in Phase 1 have equal times and in Phase 2 for each step all events have equal times then **bw** and **s** are optimal.

Notice that **s** is determined by $u_i$, $i = 1,...,n$ only. Suppose **bw** is not optimal and there is a better solution $\mathbf{bw}^{'}$ and **s** such that $\mathbf{bw}^{'} < \mathbf{bw}$. Without loss of generality, let $bw_1^{'} < bw_1$. Then $T_1(\mathbf{bw}^{'}) \ge \frac{s_1}{bw_1^{'}} > \frac{s_1}{bw_1} = T_1(\mathbf{bw})$. This implies that $T_1(\mathbf{bw}^{'}) + T_2 > T$, $\mathbf{bw}^{'}$ is not feasible and peers cannot receive all the blocks of the media within the delay bound to play media continuously. Let $s_1^{'} < s_1$ such that the events of Phase 1 have equal times $T_1(\mathbf{bw}^{'}) = \frac{s_1^{'}}{bw_1^{'}} = \frac{s_1}{bw_1} = T_1(\mathbf{bw})$. In Phase 2, the times of step *k*, are no longer equal, $T_2(\mathbf{bw}^{'}) = \frac{s_1^{'}}{u_1} \le \frac{s_1}{u_2} = T_2(\mathbf{bw})$, and we have $T_1 + T_2(\mathbf{bw}^{'}) \le T$. This implies that for $\mathbf{bw}^{'}$ a shorter size package is distributed within the delay bound $T$. We proved that if all events in Phase 1 have equal times and in Phase 2, all events in each step have equal times, **bw** and **s** are optimal. □

## APPENDIX B

TABLE III. ACIDE MODEL NOTATION LIST

| | |
|---|---|
| $T$ | Delay bound guaranteeing a continuous media playback |
| $S$ | The size of the media package distributed during $T$ |
| $\frac{S}{T}$ | Livestream ratio, the bandwidth used to distribute live media to one peer |
| $d_i$ , $u_i$ | Download, upload bandwidth of peer $i$ (maximum sustainable bandwidth throughout $T$, $u_i \le d_j$ ) |
| $u_{avg}$ | Average upload bandwidth of a cluster |
| $T_1$ , $T_2$ | Times of Phase 1 and Phase 2 respectively, $T_1 + T_2 \le T$ |
| $t_i$ | Time of step $i$ in Phase 2, $T_2 = \sum_{i=1}^{n-1} t_i$ , $t_1 = ... = t_{n-1}$ |
| $s_i$ | Size of block $i$ downloaded by peer $i$, $\sum_{i=1}^{n} s_i = S$ |
| $bw_i$ | Bandwidth allocated to peer $i$ to download $s_i$ in $T_1$ (constant during $T_1$ ), $bw_i \le d_i$ |
| $bw$ | Minimum allocated bandwidth to a cluster of $n$ peers |
| $\theta_i$ | Time to send $s_i$ sequentially in Phase 1, $T_1 = \sum_{i=1}^{n} \theta_i$ , $t_i = \frac{s_i}{bw}$ |
| $BW$ | Base station bandwidth reserved for a cluster of $n$ peers |
| $P_k$ | Livestream packages, $k = 1,...,\infty$ |

## APPENDIX C

TABLE IV. CLOSELY RELATED WORK COMPARISON TABLE

| Model \ Reference | | ACIDE | 4, 5 | 13 | 27 | 35 | 40 | 42 |
|---|---|---|---|---|---|---|---|---|
| ***Mobile Wireless Network*** | | √ | | | √ | √ | √ | √ |
| ***Live Media Streaming*** | | √ | | | √ | | √ | |
| ***High user density*** | | √ | | | | | √ | |
| ***Optimize*** | *bw* | √ | | √ | | | √ | √ |
| | *n* | √ | | | | | | |
| | *Delay* | | √ | | √ | √ | | |
| | *Energy* | | | | √ | | √ | √ |
| ***P2P*** | | √ | √ | √ | | √ | | |
| ***P2P Overlay*** | | | | √ | | √ | | |
| ***Simultaneous bidirectional connections required by a peer*** | | 1 | *n*-1 | *n*-1 | | > 1 | | |
| ***Peer $bw_i$ optimization*** | | √ | | √ | | | | |
| ***Content Delivery and D2D*** | | | | | √ | | | |
| ***Codebook Multicast and D2D*** | | | | | | | | √ |
| ***Hybrid Unicast-Multicast*** | | | | | | | √ | |
| ***Complexity*** | *Low* | √ | √ | | | √ | | |
| | *High* | | | √ | √ | | √ | √ |

## APPENDIX D

In digital communication systems, the maximum achievable transmission bandwidth is determined by the instantaneous characteristics of the communication channel. When the radio channel conditions remain relatively stable, the channel can sustain a data transmission at a constant bandwidth over a certain time interval. However, variations in propagation conditions can degrade the received signal quality and reduce transmission performance. To address these variations, transmission systems can dynamically adjust parameters such as the modulation scheme, coding rate, and symbol rate, thereby adapting the transmission bandwidth to the prevailing channel conditions. This adaptive capability enables the ACIDE model to dynamically adjust $d_i$ , $u_i$ to maintain uninterrupted livestreaming under varying channel conditions. Furthermore, the proposed Active Peer Control (APC) method dynamically adapts the delay bound to the characteristics of the underlying physical channel through the APC-DBA.

Although the physical-layer implementation of P2P communications within a cluster is beyond the scope of this paper, a brief discussion is provided to assess the feasibility and efficiency of the proposed ACIDE model. A potential physical layer realization is examined next for both a festival-based cluster and a robotaxis cluster. In both scenarios, the Phase 2 P2P communications are assumed to operate in unlicensed frequency bands, such as 5GHz or 6GHz, independently of the spectrum allocated to the base station. Within these bands, a channelized Frequency Division Multiple Access (FDMA) allocation scheme for P2P links can mitigate channel contention and reduce communication latency. Furthermore, frequency separation combined with dynamic transmit-power control can reduce intra-cluster interference, while frequency reuse can mitigate inter-cluster interference.

Given the high peer density in festival environments, QPSK modulation [53], [54] can provide reliable transmission performance, supporting data rates of several hundred kilobits per second per P2P link over coverage ranges of a few meters. Moreover, its energy efficiency enables continuous livestreaming over extended periods of time. Under these operating conditions, the unlicensed 5GHz and 6GHz bands can potentially support tens of concurrent P2P links per cluster using the ACIDE mesh or star topology. The recommended ACIDE star topology employs two concurrent unidirectional links: one for broadcasting a block to the remaining $n$ - 1 peers and another for receiving a block from one of them. This configuration maximizes spectrum utilization. The aforementioned physical-layer considerations also apply to multi-lane robotaxi clusters employing mesh or star P2P topologies. Dedicated roof-mounted antennas can improve link connectivity, and their elevated placement can extend P2P coverage within high-density, multi-lane clusters compared with festival environments.

During Phase 2, interface reconfiguration and link synchronization at the beginning of each of the $n$ – 1 sequential steps introduce additional delays, reducing livestreaming bandwidth efficiency in Phase 1. These delays depend on the underlying hardware architecture and implementation. However, specialized hardware platforms can potentially reduce the transition latency to a millisecond range.

Ultimately, the air interface, processing hardware, and propagation environment jointly determine the performance limits of the ACIDE model in terms of bandwidth efficiency, cluster coverage, and peer capacity. Nevertheless, appropriate selection of unlicensed spectrum, robust modulation and coding schemes, and deterministic channel-access mechanisms can provide sufficient design margin for practical implementations. A physical-layer realization of Phase 2 P2P communication is left for future work.

**Andrei Negulescu** received his Dipl. Ing. in Electronics and Telecommunications from University Politehnica Bucharest, and his M.A.Sc. in Electrical Engineering from Ecole Polytechnique Montréal. He is currently pursuing his Ph.D. degree in Computer Science and Engineering at Santa Clara University. His research interests include autonomous intelligent networks, distributed systems and mobile networking.

**Weijia Shang** received BS degree in computer engineering from Changsha Institute of Technology, China, and Master and Ph.D. degrees in computer engineering from Purdue University, West Lafayette, Indiana. She joined Santa Clara University in January 1994. Before that, she was on faculty of the Center for Advanced Computer Studies, University of SW Louisiana for three and half years. She received Research Initiation Award in 1991 and Career Award in 1995 form National Science Foundation. She was a Clare Boothe Luce Professor between 1994 and 2000. Her research interests include parallel processing, computer architecture, parallelizing compiler, algorithm theory and non-linear optimization.